\documentclass[
aip,
 amsmath,amssymb,
 reprint,%
]{revtex4-1}

\usepackage{graphicx}
\usepackage{dcolumn}
\usepackage{bm}

\usepackage[utf8]{inputenc}
\usepackage[T1]{fontenc}
\usepackage{mathptmx}
\usepackage{etoolbox}

\usepackage[capitalise]{cleveref}
\usepackage{upgreek}
\usepackage{booktabs}
\usepackage{mathrsfs}
\usepackage{color}

\makeatletter
\newcommand{\micron}{{\upmu\mathrm{m}}}
\newcommand{\Wcmsqd}{{\mathrm{W }\cdot\mathrm{cm}^{-2}}}
\newcommand{\kgms}{{\mathrm{kg }\cdot\mathrm{m}^{-1}\cdot}\mathrm{s}^{-1}}

\newcommand{\Rmnum}[1]{\expandafter\@slowromancap\romannumeral #1@}
\makeatother

\newcommand{\rc}[1]{\textcolor{black}{#1}}

\makeatletter
\def\@email#1#2{%
 \endgroup
 \patchcmd{\titleblock@produce}
  {\frontmatter@RRAPformat}
  {\frontmatter@RRAPformat{\produce@RRAP{*#1\href{mailto:#2}{#2}}}\frontmatter@RRAPformat}
  {}{}
}%
\makeatother
\begin{document}

\preprint{AIP/123-QED}

\title{\rc{Generation of 10 kT Axial Magnetic Fields Using Multiple Conventional Laser Beams: A Sensitivity Study  for kJ PW-Class Laser Facilities}}
\bigskip
\author{Jue Xuan Hao}
\author{Xiang Tang}%
 \affiliation{Department of Plasma Physics and Fusion Engineering, University of Science and Technology of China, Hefei 230026, China}
\author{Alexey Arefiev}

\affiliation{Department of Mechanical and Aerospace Engineering,
University of California at San Diego, La Jolla, CA 92093, USA}

\author{Robert J. Kingham}
\affiliation{%
Blackett Laboratory, Imperial College London, London SW7 2AZ, United Kingdom
}%

\author{Ping Zhu}
\affiliation{%
National Laboratory on High Power Laser and Physics, Shanghai Institute of Optics and Fine Mechanics, Chinese Academy of Sciences, Shanghai 201800, China
}%

\author{Yin Shi*}
 \affiliation{Department of Plasma Physics and Fusion Engineering, University of Science and Technology of China, Hefei 230026, China}
\email{shiyin@ustc.edu.cn}

\author{Jian Zheng}
 \affiliation{Department of Plasma Physics and Fusion Engineering, University of Science and Technology of China, Hefei 230026, China}
\affiliation{%
Collaborative Innovation Center of IFSA (CICIFSA), Shanghai Jiao Tong University, Shanghai 200240, Peoples Republic of China
}%

\date{\today}

\begin{abstract}
\rc{Strong multi-kilotesla magnetic fields have various applications in high-energy density science and laboratory astrophysics, but they are not readily available. In our previous work~[Y. Shi et al., Phys. Rev. Lett. 130, 155101 (2023)], we developed a novel approach for generating such fields using multiple conventional laser beams with a twist in the pointing direction. This method is particularly well-suited for multi-kilojoule petawatt-class laser systems like SG-II UP, which are designed with multiple linearly polarized beamlets. Utilizing three-dimensional kinetic particle-in-cell simulations, we examine critical factors for a proof-of-principle experiment, such as laser polarization, relative pulse delay, phase offset, pointing stability, and target configuration, and their impact on magnetic field generation. Our general conclusion is that the approach is very robust and can be realized under a wide range of laser parameters and plasma conditions. We also provide an in-depth analysis of the axial magnetic field configuration, azimuthal electron current, and electron and ion orbital angular momentum densities. Supported by a simple model, our analysis shows that the axial magnetic field decays due to the expansion of hot electrons.}

\end{abstract}

\maketitle

\section{\label{sec:level1}Introduction}

Strong magnetic fields are essential for research in high-energy-density (HED) science, astrophysics, and controllable nuclear fusion sciences. Plasma properties can be influenced by the presence of magnetic fields of different scales. For instance, astrophysical plasmas with a temperature and density of $10 < T < 100\ \rm{eV}$ and $10^{17} < n < 10^{19}\ \rm{cm}^{-3}~$~\cite{Plechaty2010,Albertazzi2014,Schaeffer2017,Byvank2017,Matsuo2019}, respectively, can be magnetized with a magnetic field strength of about 10 T. Relativistic magnetic reconnection can be observed in such plasmas when the magnetic field strength reaches 1 kT ~\cite{Yi2018,Ping2021,Raymond2018,Law2020,Gu2016}, since the Alfven velocity becomes comparable to the speed of light. In contrast to that, the HED plasmas found in the cores of stars and planets, and in the fuel of inertial confinement fusion, have much higher temperatures and densities, exceeding $1\ {\rm{keV}}$ and $10^{21}\ {\rm{cm}}^{-3}$, respectively. For these plasmas, a magnetic field strength of around 100 T can benefit electron transport when the cyclotron frequency is equivalent to or bigger than the collisional frequency. Additionally, a magnetic field strength of about 1 kT can be used to control relativistic electron beams, improving high-energy ion acceleration~\cite{Arefiev2016,Yao2024}. When the magnetic field strength exceeds 10 kT, the laser can propagate in the magnetized plasma with electron density above the critical density as a whistler mode~\cite{Luan2016,Sano2017,Li2020,LiK2023}. The extremely small electron Larmor radius in magnetic fields above 10 kT leads to characteristic quantum-mechanical phenomena~\cite{Liu2019,Higuchi2018,Liseykina2021}. The energy density of a magnetic field is given by $\varepsilon_B = B^2/(2\mu_0) $, which can be expressed as $\varepsilon_B~[{\rm{J\cdot m}}^{-3}] \approx 4 \times 10^{11} B^2~[\rm{kT}]$. When the magnetic field strength exceeds 1 kT, we can study the properties and dynamics of matter under extreme conditions where the magnetic field energy density is greater than $10^{11}~\rm{J\cdot m}^{-3}$.
This may bring new  opportunities to the HED science. 

The generation of strong magnetic fields in plasmas is of great interest due to its potential for a wide range of applications~\cite{Herlach2003}, including magnetically enhanced fast-ignition fusion~\cite{Strozzi2012}, generation of collisionless shocks in magnetized plasmas~\cite{Sagdeev1966,Yao2021}, magnetically assisted ion acceleration~\cite{Kuznetsov2001,Fukuda2009,Nakamura2010,Bulanov2015,Park2019,Gong2017}, and magnetic field reconnection research~\cite{Gu2021}. In order to investigate the impact of strong magnetic fields, various approaches have been proposed and developed to generate fields over 100 T in laboratory environment. These include the use of pulsed-power devices~\cite{Albertazzi2013,Shapovalov2019,Hu2020}, self-generated magnetic fields~\cite{Mason1998,Roth2016,Wagner2004,Yang2023,Zosa2022}, magnetic flux compression~\cite{Nakamura2018,Sefkow2014,Gotchev2009,Moody2021}, and laser-driven coils~\cite{Santos2015,Kochetkov2022,Vlachos2024}. The corresponding measurement techniques of the magnetic field have also made significant strides in development~\cite{Russell2023,Zhu2022}. The magnetic flux compression technique has been a widely used method for generating high magnetic fields in laboratory experiments. One advantage of this technique is the use of non-destructive, single-turn coils driven by pulsed-power devices to generate magnetic fields of several tens of tesla for long durations ($ > 1~\rm{\upmu s} $) and large volumes ($ > 1~\rm{mm}^3 $) ~\cite{Albertazzi2013,Shapovalov2019,Hu2020}. By utilizing these magnetic fields as a seed field, a strong magnetic field over 100 T can be generated using the flux compression method. 
Alternatively,  the scheme based on laser-driven coil can utilize high-power lasers with an intensity of approximately $10^{15}~\Wcmsqd$ to generate a magnetic field exceeding 100 T without requiring compression and a seed magnetic field. However, the underlying mechanism of how the laser-driven coil produces such a strong magnetic field is not fully understood, and its effectiveness remains a topic of debate. In a recent study, Peebles et al.~\cite{Peebles2022} conducted extensive experiments to assess the potential of laser-driven coils in generating strong magnetic fields. The authors used different types of laser-driven coils to generate magnetic fields and employed various diagnostic techniques such as B-dot probing, Faraday rotation, and proton radiography~\cite{Liao2016} to measure the fields. Their conclusion is that the laser-driven coils cannot create quasi-static kilotesla-level fields as claimed~\cite{Peebles2022}.

At the same time, high-intensity laser systems have become increasingly important in strong magnetic field generation by inducing strong electric currents. With the invention of the Chirped Pulse Amplification (CPA)~\cite{Mourou1985} technique, laser pulses are amplified to possess an ultra-high energy density. A laser pulse with high-intensity is an excellent driver for strong electrical currents suitable for generation of extremely strong magnetic fields. During the interaction between a high-intensity laser and an overdense plasma, the self-generated magnetic field on the plasma surface can be close in strength to the oscillating laser magnetic field~\cite{Tatarakis2002}. Though such high amplitude magnetic fields are not in the bulk and have a complex topology, they can have an effect on generation of hot electrons~\cite{LiXX2023}.

{Generation of strong magnetic fields in a large volume requires the use of laser beams with both high-intensity and high energy. Assuming the energy conversion from the laser into magnetic fields is 0.5\%,  a laser energy as high as 80~J is required to generate a 1~kT field with a volume of $(100~\micron )^3$. If the volume of the laser beam at focus is the same as the volume of the magnetic fields and the energy conversion is around 1\%,  we can expect that the strength of the generated magnetic field is going to be around $B \approx 0.1  B_L~$~\cite{Stark2016}, where $B_L$ is the peak amplitude of the laser magnetic field.} 
There has been a significant increase in the number of laser facilities around the world that are capable of producing peak power at the petawatt (PW) level~\cite{Danson2019}. Meanwhile, the beam energy of several systems can be in the multi-kJ range. Such multi-kJ PW-class laser systems as LFEX~\cite{Kawanaka2008}, NIF ARC~\cite{Crane2010}, and Petal~\cite{Batani2014}, composed of multiple linearly polarized (LP) beamlets, allow experimental investigation of magnetic field generation in bulk plasma within relativistic regime by delivering the highest energy within picoseconds. The multi-beamlet configuration is not only an essential feature of the laser system design, but also the key to advanced laser-plasma interaction regimes~\cite{Morace2019,Yao2024}. 
The SG-\Rmnum{2} UP facility~\cite{Zhu2018}, which includes a picosecond (ps) petawatt laser and tens of kilojoules (kJ) nanosecond (ns) lasers, is now being upgraded to a larger scale laser physics platform with multiple kJ-class ps laser beams and hundreds of kJ ns lasers. With the increased number of laser beams, the upcoming extended platform will provide the multi-beamlet experimental capacity of kJ-class ps petawatt lasers. 

One particular mechanism for producing self-generated magnetic fields is through the inverse Faraday effect (IFE), where an axial magnetic field is spontaneously generated when the laser transfers angular momentum (AM) to plasma electrons. There exist various methods for generating axial magnetic fields through AM transfer. In the early days, the IFE was mostly observed when circularly polarized (CP) radiation propagated through an unmagnetized plasma, resulting in the generation of a quasistatic axial magnetic field~\cite{Haines2001}, yet the effect is not exclusive to a CP beam carrying spin angular momentum (SAM) but also applicable to a vortex beam carrying orbital angular momentum (OAM). It is now well-known that helical wavefronts can be represented in a basis set of orthogonal Laguerre-Gaussian (LG) modes and that each LG mode is associated with a well-defined state of photon OAM~\cite{Allen1992}. The magnetic field generation using LG beams has been investigated theoretically and numerically~\cite{Ali2010, Nuter2020, Longman2021}. 

While the IFE is widely studied, the optical technology and elements required for producing CP or LG laser beams from conventional LP beams at high power can be costly, complex, and fragile. As of now, the generation of adequately strong and precisely controllable macroscopic fields at laser facilities employed in HED physics research~\cite{Danson2019, Li2022, Zhu2018} remains a significant challenge. Recently, we have proposed a novel multi-beam approach for AM transfer to a plasma, leading to subsequent strong magnetic field generation~\cite{Shi2023}. The approach involves a spatial arrangement of four conventional laser beams, depicted in \cref{Scheme}, that is inspired by the multi-beam design of the PW-class kJ laser systems. The key aspect of our design is its ability to overcome the challenge faced by conventional LP Gaussian lasers, which is that they do not possess intrinsic AM and thus cannot achieve the IFE. By employing the specific spatial arrangement of laser beams, we enable the transfer of ensemble AM to the plasma, inducing strong rotating currents, that lead to efficient generation of a strong axial magnetic field. This new multi-beam approach shows promising potential for enhancing magnetic field generation in laser-driven plasma systems.

\rc{Building on our previous work~\cite{Shi2023}, this paper provides an in-depth analysis of the multi-beam approach. We begin by examining the angular momentum (AM) carried by several conventional laser beams in \cref{Sec-2} to frame the study in the context of AM transfer. In \cref{Sec-3}, we present the results of a 3D particle-in-cell (PIC) simulation for four laser beams with twisted pointing directions and a target containing a preplasma-like density ramp. A simple model of the axial magnetic field decay based on the expansion of hot electrons is provided.  This section essentially reviews the key findings from Ref.~[\onlinecite{Shi2023}], albeit for a different target, setting the stage for the subsequent discussion in \cref{Sec-4}. In \cref{Sec-4}, we examine various factors likely to impact field generation in actual experiments, including laser polarization, relative pulse delay, phase offset, laser pointing stability, and target configuration. \Cref{Sec-5} compares our scheme with other methods that use circularly polarized (CP) and Laguerre-Gaussian (LG) laser beams to drive the magnetic field. Finally, \cref{Sec-6} summarizes the main results of this work.}


\section{\label{Sec-2}Angular momentum carried by laser beams}
It is well known that paraxial optical beams carry three distinct types of AM~\cite{Bliokh2015}. 
They are SAM for CP beams, intrinsic OAM (IOAM) for LG beams with helical wavefronts and extrinsic OAM (EOAM) for normal beams propagating at a distance from the coordinate origin. The SAM is determined by the right- or left-hand CP. The IOAM is determined by the twist index of the LG beams. Both the SAM of a CP beam and the IOAM of a LG beam can be used for axial magnetic field generation via IFE~\cite{Haines2001,Ali2010, Nuter2020, Longman2021}. For EOAM, we can understand it from the viewpoint of photons carrying linear momentum $\bm{p}$ at position $\bm{r}$. The EOAM can be calculated as ${\bm{L}}^{\rm ext}=\bm{r\times p}$. 

\begin{figure*}[ht]
    \centering
    \includegraphics[width=0.9\textwidth]{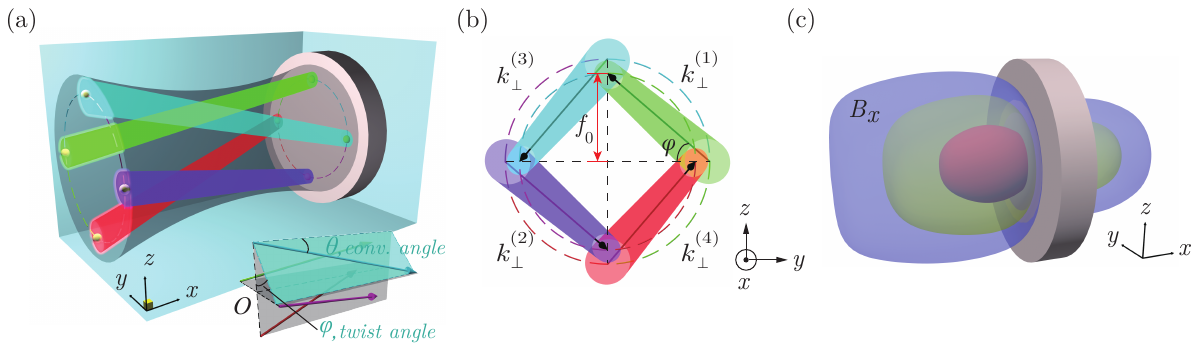}
    \caption{(a) 3D schematic of the target and four regular laser beams with twisted pointing directions carrying AM collectively. The colored columns indicate the beams and the corresponding dash lines show the way that the twist is induced by the beams. (b) 2D projection of the beams onto the $yz$ plane, showing the beginning and the end in the emitter plane and the focal plane, respectively. Each arrow is the component of the wave-vector (in a corresponding colored beam) transverse to the global direction ($x$-axis) of propagation of the ensemble of beams. The parameters can be found in \cref{LPI par}. (c) Illustration of the axial magnetic field $B_x$ after the lasers have left the simulation box ($t=20 ~\rm fs$). The blue, yellow, and red isosurfaces, from the outside to the inside, represent increasing magnetic field strength. The axial magnetic field profile is quantified in \cref{bx3d&r}.}
    \label{Scheme}
\end{figure*}

In order to take advantage of EOAM like SAM or IOAM in the IFE, we have proposed a setup that involves four laser beams with twisted pointing directions~\cite{Shi2023} that is schematically shown in \cref{Scheme}. Each laser beam is represented by a colored cylindrical cone, with its direction determined by a wave vector $\bm{k}_i$, where $i$ is the index numbering the beam. The photon momentum in the $i$-th beam is $\bm{p}_{i} = \hbar \bm{k}_{i}$. For simplicity, consider two beams, the green and purple ones, with $\bm{k}_{(1,2)} = (k_x, k^{(1,2)}_{\perp}, 0)$, intersecting the $(y,z)$-plane at $z_{(1,2)} = \pm f_0$ and $y_{(1,2)} = 0$, respectively, where $f_0$ is the beam offset. In this case, the axial AM of a given photon is given by $[\bm{r} \times \bm{p}]_x$, where $\bm{r}$ is the position vector and $\bm{p}$ is the photon momentum. As a result, the total AM of the two beams is approximately 
\begin{equation} \label{eq: 1}
    L_x \approx - N \hbar \left( k_{\perp}^{(1)} - k_{\perp}^{(2)} \right) f_0 ,
\end{equation}
where $N$ is the number of photons in one beam. If the two beams have the same tilt, then $k^{(1)}_{\perp} = k^{(2)}_{\perp}$ and, as a result, $L_x \approx 0$. If the tilt of the second beam is opposite to that of the first, then $k^{(2)}_{\perp} = -k^{(1)}_{\perp}$. As a result, the total AM no longer vanishes and it is roughly given by $L_x \approx - 2 N \hbar k_{\perp}^{(1)} f_0$. The total AM will double when a pair of such lasers with an offset in the $y$ direction is added, showing that appropriately arranged beams can carry AM even though individual photons in each beam have no intrinsic AM~\cite{Bliokh2015}. This observation bears similarities to studies involving $\gamma$-ray beams carrying OAM~\cite{Liu2016, Chen2018, Chen2019}, where a population of photons with a twisted distribution of momentum $\bm{p}$ is generated.
An alternative method is to calculate the AM of electromagnetic field as 
\begin{equation}
    \bm{L} = \varepsilon_0 \int \left( \bm{r} \times \left[ \bm{E} \times \bm{B} \right] \right) d^3 {r}.
\end{equation}
where $\varepsilon_0$ is the dielectric permittivity and $\bm{E}$ and $\bm{B}$ are the electric and magnetic fields, respectively. Note that $[\bm{E} \times \bm{B}]$ matches the direction of the Poynting vector. In a given laser beam, the dominant component of the Poynting vector is directed along the axis of the beam. Therefore, four laser beams with a twist in the pointing direction (shown in \cref{Scheme}) can possess net AM. 

\section{\label{Sec-3} Simulation results for beams with twisted pointing directions} 

In this section, we present a self-consistent analysis performed using three-dimensional (3D) particle-in-cell (PIC) simulations with the open-source relativistic PIC code EPOCH~\cite{Arber2015}. \rc{The key findings align with those first reported in Ref.~[\onlinecite{Shi2023}]. The overlap between parts of this section and Ref.~[\onlinecite{Shi2023}] is intentional, serving to set the stage for the discussion in \cref{Sec-4}. }

As explained in \cref{Sec-2}, our goal is to take advantage of the multi-beam configuration available at some state-of-the-art PW-class laser systems. We consider a laser system that provides four identical LP Gaussian laser beams that have no intrinsic AM. The peak intensity of each laser beam is $I_0 = 8.0 \times 10^{19}~\Wcmsqd$, which  corresponds to a normalized electric field amplitude $a_{0}= 8.0$. The normalization of the electric field is defined as $a_0 = eE_0/m_e \omega c$, where $E_0$ is the peak electric field amplitude. Where $c$ is the speed of light, $\omega$ is the center frequency of the laser beam, and $e$ and $m_e$ are the electron charge and mass, respectively. The pulse duration, $\tau_g=600$ fs, is the same for each beam  (the temporal envelope of the electric field is Gaussian). The laser wavelength is $\lambda =1.053~\micron$ and the beam waist radius is $w_0 =6.0~\micron$.

\begin{table*}
    \centering
    \begin{tabular}{ ll }
        \toprule
        \multicolumn{2}{l}{\textbf{Parameters of four laser beams}}\\
        \midrule
        Peak intensity & $I_0 = 8.0 \times 10^{19}~\Wcmsqd$ \\
        
        Normalized field amplitude & $a_{0}= 8.0$ \\
        
        Wavelength & $\lambda = 1.053~\micron$ \\
        
        Focal spot size ($1/\mathrm{e}$ electric field) & $w_0 = 6.0 ~\micron$ \\
        
        Pulse duration ({Gaussian electric field}) & $\tau_g=600$ fs\\
        
        The global direction of propagation of the ensemble of beams & $+x$ \\
        
        Linear polarization in the emitter plane & $+y$ \\
        
        Location of the emitter plane & $x_{e} = -20.0~\micron$\\
        
        Location of the focal plane & $x_{f} = -5.0~\micron$\\
        
        Beam offset in the focal plane & $f_{0} = 11.0~\micron$\\
        
        The polar and azimuth angle for convergence and twist &$\theta = 0.27\pi$, $\varphi = 0.28\pi$ \\
        \midrule
        \multicolumn{2}{l}{\textbf{Other parameters}}\\
        \midrule
        Foil thickness & $x\in [0,3] ~\micron$\\
        
        Preplasma thickness & $x\in [-5,0] ~\micron$\\
        
        Modulation mode of preplasma & $+x$ exponential\\
        
        Electron density & $n_e = 50.0\;n_{c}$ \\  
        
        Ion (C$^{6+}$) density & $n_i = 50.0/6\;n_{c}$ \\  
        
        Simulation box & $(40~\micron)^3$\\  
        
        Spatial resolution & 25 cells/$\micron$ \\
        
        Macroparticles per cell &  4 \\
        
        Location of the front surface of the foil & $x = 0~\micron$ \\
        
        Time when the laser beams leave the simulation box & $t = 0$~fs \\
        \bottomrule
    \end{tabular}
    \caption{3D PIC simulation parameters. $n_{c} = 1.0\times 10^{21}$~cm$^{-3}$ is the critical density for the considered laser wavelength. The initial temperature is set to zero. The electron to ion mass ratio is $1/(1836 \times 12)$.}
    \label{LPI par}
\end{table*}

\begin{figure*}[htb]
    \centering
    \includegraphics[width=0.85\textwidth]{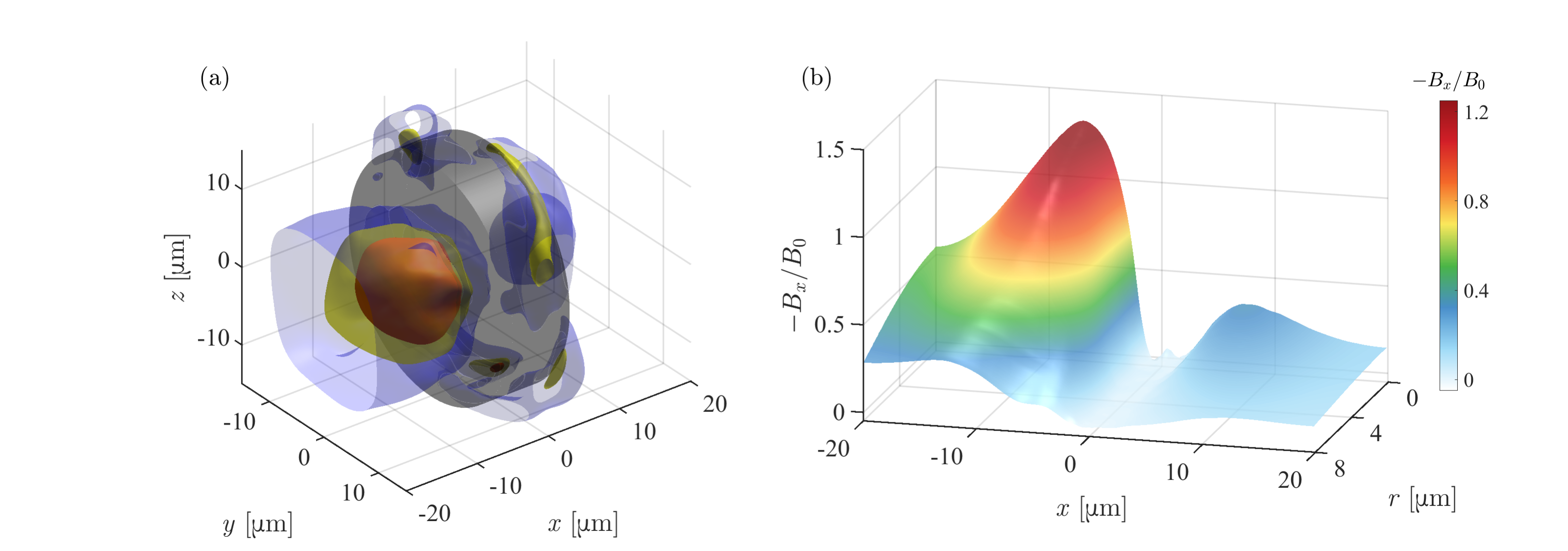}
    \caption{Distribution of the axial magnetic field at $t = 20$~fs. (a) Volumetric isocontours of axial magnetic field component for $B_x = -0.2$, $-0.55$, $-0.8~B_0$ (purple, yellow, red). (b) Distribution of the axial magnetic field in the $(x,r)$-plane. Azimuthal averaging is performed.}
    \label{bx3d&r}
\end{figure*}

\begin{figure*}[htb]
    \centering
    \includegraphics[width=0.85\textwidth]{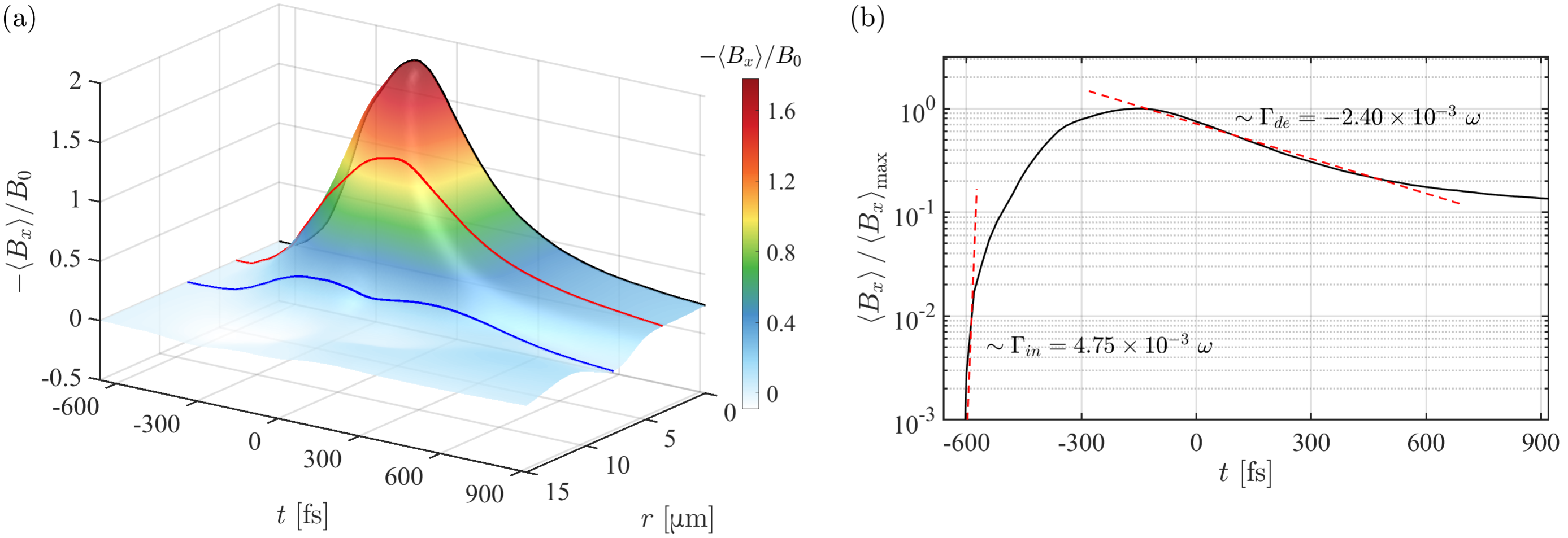}
    \caption{Temporal evolution of the axial magnetic field. (a) The distribution of the axial magnetic field in the $(t, r)$-plane, with slices highlighted at radial distance $r=0~\rm{(black)}$, $4~\rm{(red)}$, $8~\rm{(blue)}~\micron$. Longitudinal and azimuthal averaging is performed. (b) Axial magnetic field for $r=0~\micron$, where $\max\left\langle B_x\right\rangle = 1.78~B_0$. $t=0~\rm fs$ is defined as the moment when the lasers leave the simulation space completely.}
    \label{bxbar_t}
\end{figure*}

The orientation of the four laser beams is set according to \cref{Scheme} (a), where the lasers are represented by colored conical cylinders. For each beam, the axis represents the propagation direction and the transverse size represents the beam radius. The lasers enter the simulation domain from the left boundary of the simulation box ($x_e = - 20~\micron$) that we refer to as the emitter plane. They interact with the plasma at the focal plane located at $x_f = - 5~\micron$.
The transverse electric field of each beam in the $(y,z)$-plane is set to be directed along the $y$-axis. \Cref{Scheme} (b) shows projections of all four beams onto the $(y,z)$-plane. The beams are set up such that the intersection points of the beam axes with the $(y,z)$-plane for a given longitudinal position $x$ form vertices of a square. This square formed by the intersection points rotates and shrinks as the beams propagate from the emitter plane to the target. To make it more evident that the intersection points get closer to each other as the beams approach the target, we also plotted two circles that go through the intersection points in the emitter plane and the focal plane. The circle in the focal plane is visibly smaller.
The radius of the circle in the focal plane is $f_0$. We call it the beam offset. The twist degree of four laser beams is controlled by the azimuthal angle $\varphi=0.28\pi$. By definition, there is no twist for $\varphi = 0$. The angle between the axis of every beam and the focal plane (in the plane formed by the beam axis and the $x$-axis) is $\theta = \arctan(S/D)=0.27\pi$, which is the polar angle characterizing the beam convergence, where $D$ is the distance between the emitter plane and the focal plane and $S$ is the transverse shift of the beam axes in the two-dimensional projection plane.

We assume that the target is a fully ionized carbon plasma with an exponential longitudinal density profile to mimic a preplasma. The initial electron density is set to $n(x) = 0.05n_e{\rm{exp}}\{(x~[\micron]+5)/1.67\}$, where $n_e=50\;n_{c}$ is the electron density of the foil whose thickness is $3~\micron$  behind the preplasma and $n_{c} = 1.0\times 10^{21}$~cm$^{-3}$ is the critical density corresponding to a laser wavelength $\lambda = 1.053~\micron$. Both electron and ion populations are 
initially cold ($T$ = 0). The front surface of the foil is at $x=0$ and the rear surface is at $x=3~\micron$. 
Using this profile, the plasma density ramps up from $3\;n_{c}$ to $50\;n_{c}$ over $5~\micron$, which increases the interaction volume between the laser beams and the plasma. The simulation box size is $(40~\micron)^3$ with grid cell sizes of $(40~\rm nm)^3$. We use four macro-particles per cell and have open boundaries throughout. \rc{Note that our resolution and the number of macro-particles per cell are comparable to what has been used in other studies of dense laser-irradiated targets and plasma-generated magnetic fields (e.g. see Ref.~[\onlinecite{weichman2020protons}]).} \Cref{LPI par} gives detailed parameters of the 3D PIC simulation presented here. It must be pointed out that our setup differs from that used in Ref.~\onlinecite{Shi2023}, where we employed nanowires rather than a preplasma and the laser wavelength was set to $0.8~\micron$. We define the moment when the laser beams leave the simulation domain following their reflection off the target as $t = 0$~fs.
In the described simulation, we observe generation and gradual evolution of an axial magnetic field. The illustration of the axial magnetic field $B_x$ after the lasers have left the simulation box ($t = 20$~fs) is shown in \cref{Scheme} (c). The blue, yellow, and red isosurfaces, from the outside to the inside, represent increasing magnetic field strength.

\begin{figure*}[htb]
    \centering
    \includegraphics[width=0.8\textwidth]{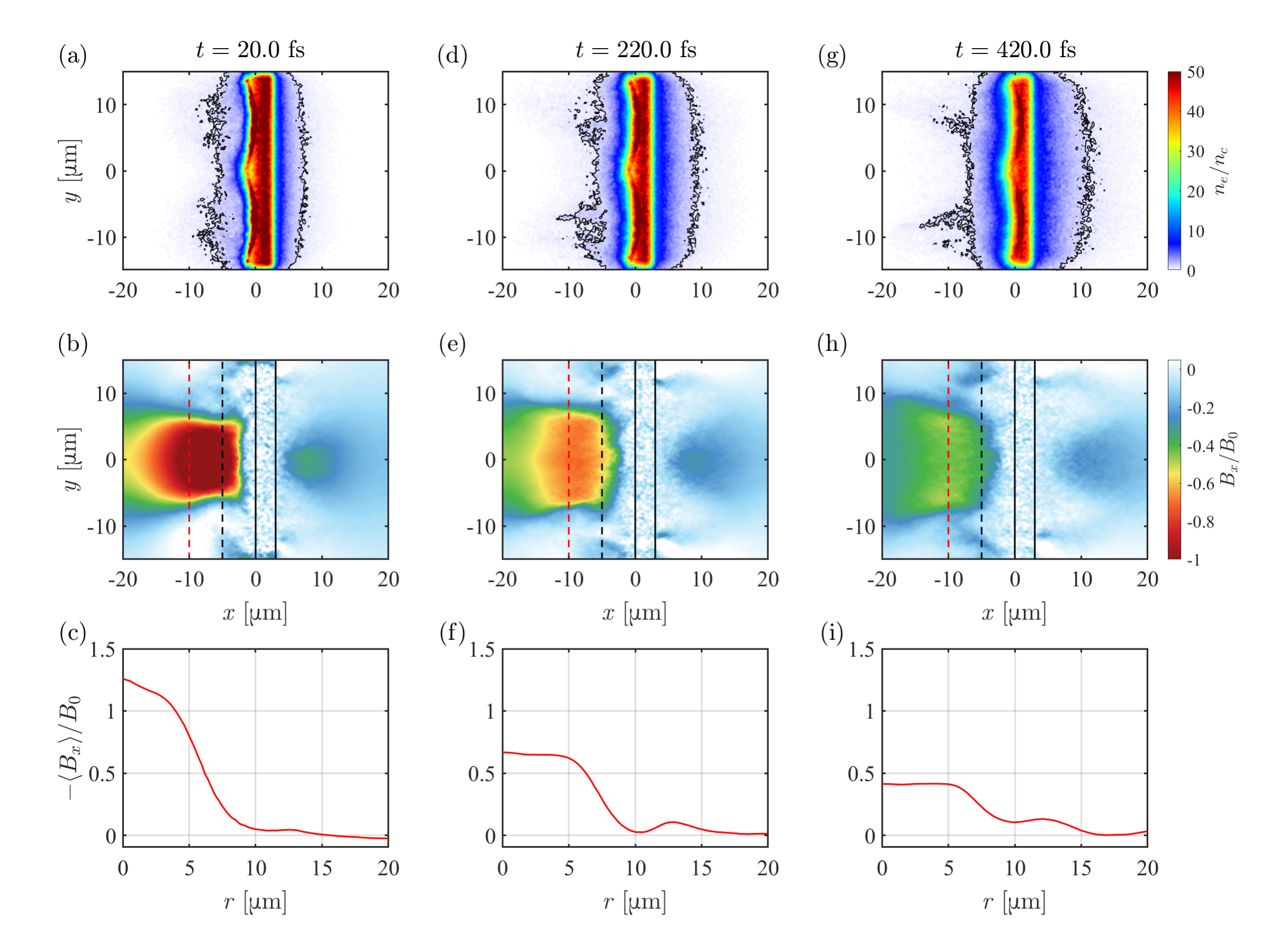}
    \caption{Simulation results at three different times, $t = 20.0$, 220.0, and 420.0 fs (from left to right). (a), (d), and (g) Longitudinal slices of electron density, where the density is normalized to the critical density and the black solid lines mark $n_e = n_c$. (b), (e), and (h) Longitudinal slices of the axial magnetic field, where the field is averaged over 20 fs and normalized to $B_0=2\pi m_ec/|e|\lambda=10.0~\rm kT$; the two black solid lines respectively represent the front ($x=0~ \micron$) and rear ($x=3~\micron$) surfaces of the solid part of the target; the front surfaces ($x=-5~ \micron$) of exponentially modulated preplasma are represented by the black dashed lines; the red and black dashed lines show the boundaries of the lineout region. (c), (f), and (i) Radial lineouts of the axial magnetic field, where the field is averaged azimuthally, temporally, and longitudinally (over the lineout region marked with red and black dashed lines in the middle row). Laser-plasma parameters are given in \cref{LPI par}.}
    \label{LP_0_neBx}
\end{figure*}

\subsection{Magnetic field distribution}
We observe generation of a strong axial magnetic field in the simulation when using beams with twisted pointing directions.
\Cref{bx3d&r} (a) displays the axial magnetic field for $\varphi = 0.28 \pi$ at $t = 20$~fs. The strength of the peak longitudinal magnetic field exceeds 10~kT. The volume occupied by the field stronger than 2~kT is approximately $10^4~\micron^3$. The three volumetric isocontours indicate $B_x/B_0 = -0.2$, $-0.55$, and $-0.8$, where $B_0=2\pi m_ec/|e|\lambda=10.0~\rm kT$. To provide a clearer $B_x$ profile, the values are temporally averaged over a 20~fs interval and spatially smoothed using a box with a stencil size of $(0.4~\micron)^3$. 
Due to the approximately axisymmetric distribution of the magnetic field, for simplicity, we perform azimuthal averaging that yields the magnetic field strength distribution in the $(x, r)$-plane that is shown in \cref{bx3d&r} (b). The magnetic field reaches its peak strength of $1.4~B_0$ on axis at $x=-10~\micron$.

\rc{A strong axial magnetic field is present not only in front of the target, but also behind it. 
The magnetic field strength and its variation gradient are generally larger in front of the target. Although the magnetic field  behind the target is weaker than the magnetic field at the front, its strength can still reach thousands of Tesla. The magnetic field behind the target is generated by hot electrons carrying OAM that go through the target and exit at the rear side. The presence of the longitudinal field behind the target can be viewed as a confirmation that our scheme does indeed produce OAM-bearing electrons, since the lasers are unable to directly access this region.}

To study the temporal evolution of the axial magnetic field, we perform longitudinal and azimuthal averaging over a cylindrical region of length $L$:
 $\left< B_x\right>(r)=1/(2\pi L)\int_0^{2\pi}\int_0^{L}B_x(r,\phi,x)dxd\phi$. 
We choose  $L = 5~\micron$ according to the simulation result and show the distribution of $\left< B_x\right>(r, t)$ in \cref{bxbar_t} (a). We can see in \cref{bxbar_t} (b) that the magnetic field between $t = -600$~fs and $t =-500$~fs has the maximum growth rate $\Gamma_{in}=4.75\times10^{-3}\omega$, where  $\omega$ is the laser frequency corresponding to the laser wavelength $\lambda = 1.053~\micron$. As the hot electrons expand, the axial magnetic fields decay. We find that the characteristic decay rate is $\Gamma_{de}=-2.4\times10^{-3}\omega$. 
The duration of the axial magnetic field is around $\tau_B \approx 1 $~ps, which is in the same order of magnitude as the magnetic field generated by the IFE~\cite{Longman2021}.

\Cref{LP_0_neBx} shows the axial magnetic field distribution at three different times: panels (a)~-~(c) correspond to $t = 20.0$~fs, panels~(d)~-~(f) correspond to $t = 220.0$~fs, and panels (g)~-~(i) correspond to $t = 420.0~\rm fs$. 
\Cref{LP_0_neBx} (a), (d), and (g) give the electron density normalized to the critical electron density $n_c$. The black contours denote $n_e = n_c$. The simulation confirms that the laser beams are reflected rather than being transmitted through the foil. 
Panels (b), (e), and (h) in \cref{LP_0_neBx} are transverse slices of the axial magnetic field (averaged over 20 fs output dumps and normalized to $B_0=2\pi m_ec/|e|\lambda=10.0~\rm kT$). The two black solid lines represent the front ($x=0~ \micron$) and the back ($x=3~\micron$) of the foil, respectively. The front ($x=-5~ \micron$) of the exponentially modulated preplasma is represented by the black dashed lines. We find that the magnetic field region moves axially outward. Panels (c), (f), and (i) in \cref{LP_0_neBx} show radial lineouts of the axial magnetic field averaged azimuthally, temporally (over 20 fs), and longitudinally between the red and black dashed lines in \cref{LP_0_neBx} (b), (e), and (h). The results show that the magnetic field expands radially outwards, which leads to the reduction of its peak strength. 

\begin{figure*}[htb]
    \centering
    \includegraphics[width=0.8\textwidth]{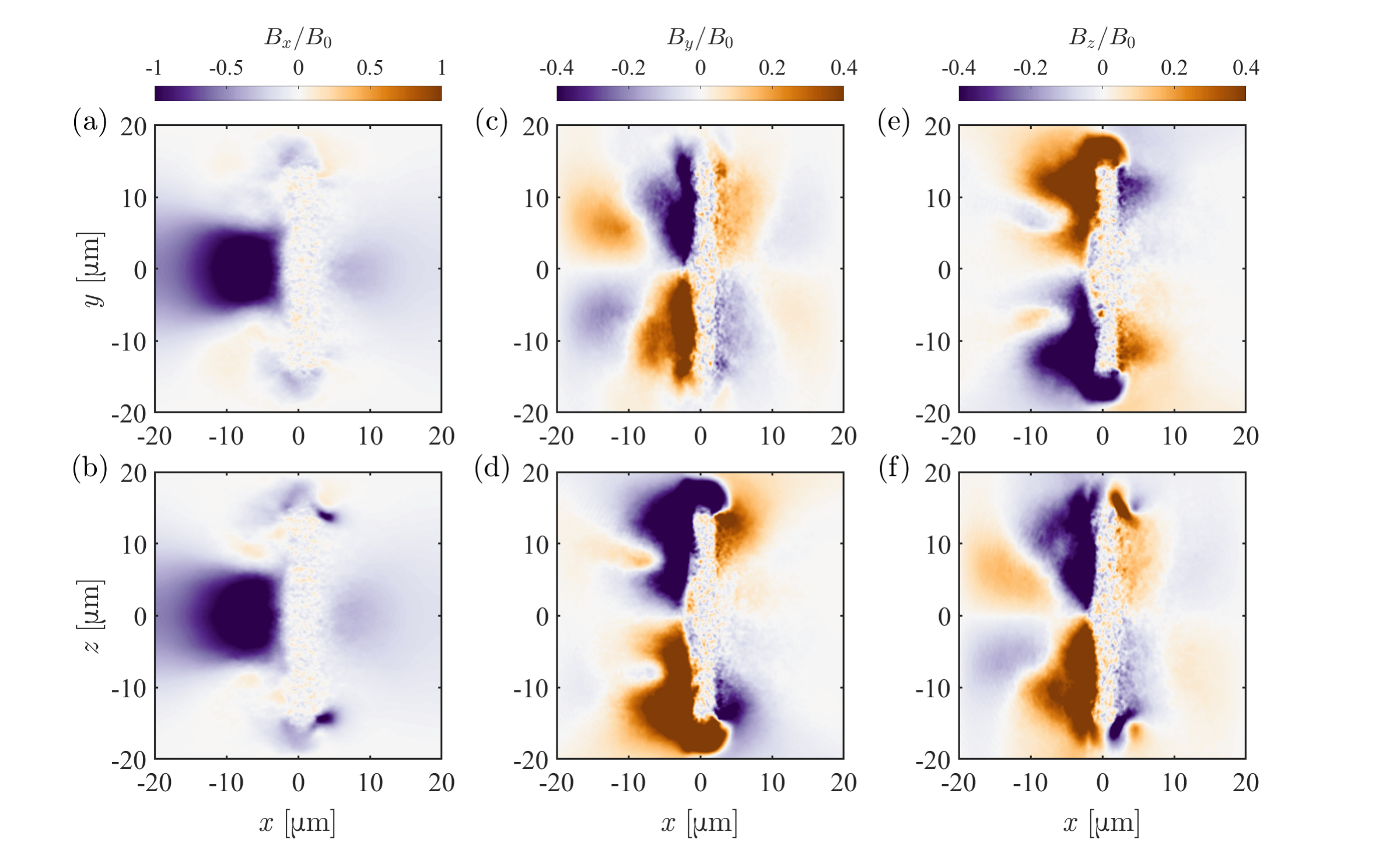}
    \caption{Two-dimensional longitudinal slices at $t = 20$~fs of all three components of the time-averaged (averaged over 20 fs output dumps) magnetic field from the 3D PIC simulation with the parameters listed in \cref{LPI par}. The fields are normalized to $B_0=2\pi m_ec/|e|\lambda=10.0~\rm kT$.  $B_x$, $B_y$, and $B_z$ are shown in $(x,y)$-plane (top row) and $(x,z)$-plane (bottom row), respectively. Note that the color scale for $B_x$ is different from that for $B_y$ and $B_z$.}
    \label{bxyz2d}
\end{figure*}

\begin{figure*}[htb]
    \centering
    \includegraphics[width=0.95\textwidth]{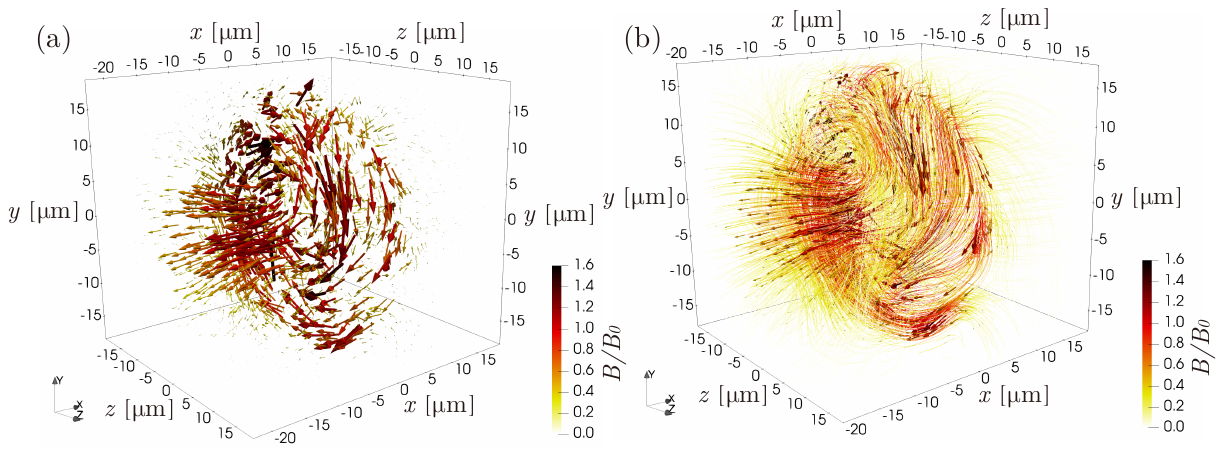}
    \caption{\rc{Three-dimensional magnetic vector field and magnetic field line structure at $t = 20$~fs. (a) The vector arrows of magnetic field, where both the length and color of each vector represent the relative field strength. (b) Magnetic field lines reconstructed from the vector field. Areas with a higher density of magnetic field lines have a higher magnetic field strength, which can be visually observed from the color. The videos in the Supplemental Material show different points of view of the distribution.}}
    \label{B_line-3D_vv}
\end{figure*}

\rc{Though our main focus is on the axial magnetic field, we provide longitudinal slices of all three components of the generated magnetic field at $t = 20$~fs. In \cref{bxyz2d}, $B_x$, $B_y$, and $B_z$ are shown in $(x,y)$-plane (top row) and $(x,z)$-plane (bottom row), respectively. These slices provide important reference values for the subsequent discussion on the dependence of the magnetic field on the twist angle. These results also confirm that the axial magnetic field is indeed very axisymmetric in our scheme. 
It is worth noting that in the region of peak axial magnetic field, the transverse components are generally an order of magnitude smaller in strength.  Furthermore, these transverse fields reach their peak values away from the central core of the axial magnetic field.
These transverse components are particularly weak near the axis. 
The described behavior can be clearly seen in \cref{B_line-3D_vv} that provides the entire vector field structure and field lines in 3D at $t = 20$~fs. The videos in the Supplemental Material show different points of view of the distribution. We can conclude that the magnetic field is axial only in the central region. }

\rc{\subsection{Magnetic field decay model based on hot electron expansion}}

\rc{To examine the efficiency of the magnetic field generation, we consider a box in front of the target with $x\in(-15,-5)~\micron$ and $y$, $z\in(-10,10)~\micron$. We find that the energy in the magnetic field is $U_B = \int B_{x}^{2}/(2 \mu_0) dV \approx 1.6$~J. The total kinetic energy of plasma electrons within the same box is roughly ten times higher, $U_{e}\approx 20$~J.
The total incident energy in the four laser pulses is $U_{\rm L} \approx 217$~J. We thus conclude that the energy conversion efficiency from the lasers to hot electrons and from hot electrons to the plasma magnetic field are both around 10\%. {This overall conversion efficiency of $\sim 1\%$ is similar to the conversion efficiency for the Biermann battery magnetic fields experimentally generated in laser-solid interactions~\cite{griffmcmahon2023measurements}.}
However, it is two orders of magnitude higher than the conversion efficiency for a laser-driven coil in Ref.~\onlinecite{Callejo2022}. This demonstrates the effectiveness of our multi-beam approach in efficiently converting laser energy into a strong magnetic field.}

\begin{figure}[htb]
    \centering
    \includegraphics[width=0.9\columnwidth]{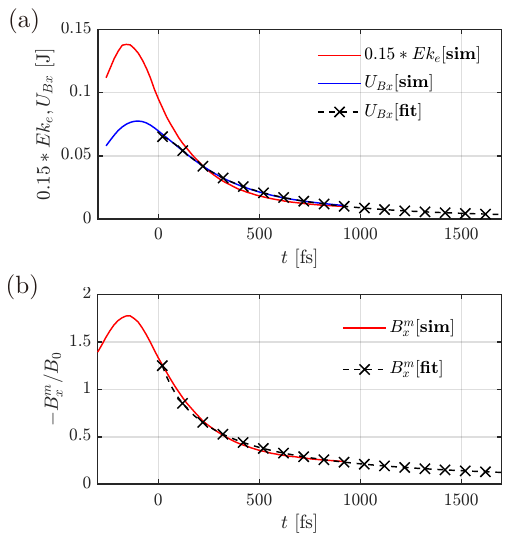}
    \caption{\rc{(a) Temporal evolution of electron kinetic energy and axial magnetic field energy within a rectangular volume of length $a=17.9~\micron$ ($x\in[-19.9,-2.0]~\micron$), width and height $b$, $c=20~\micron$ ($y$, $z\in[-10.0,10.0]~\micron$). The time $t=0~\rm fs$ is defined as the moment when the lasers completely exit the simulation domain. The red solid line represents the simulated kinetic energy of hot electrons within the region. The blue solid line depicts the simulated magnetic field energy, while the black dashed line with cross marks indicates the fit result of our model. (b) Temporal evolution of the peak axial magnetic field strength from simulation, alongside the results of $B_x^m$ derived from the fitting formula~\cref{Eq:Bscaling}. The results of the fitting model in (a) and (b) are both shown for $t$ in the range from $t = 0$~fs to $t = 1700$~fs.}}
    \label{ebEth}
\end{figure}

\rc{We examined the kinetic electron energy \( E_k \) and the magnetic field energy \( U_{Bx} \) in a region in front of the target, defined by \( x \in (-20, -2)~\micron \) and \( y, z \in (-10, 10)~\micron \). The time evolution of these two energies is shown as red and blue lines in \cref{ebEth}(a). The decay trends of both energies become nearly identical shortly after the laser pulses leave the simulation box (this occurs at \( t = 0 \) fs). This suggests that the observed decay of the magnetic field is primarily caused by the expansion of hot electrons, whose azimuthal current sustains the field. The expansion of hot electrons is influenced by several factors, including ion dynamics and potentially self-generated axial magnetic fields. A detailed study of the hot electron expansion will be addressed in future work.}

\rc{To give a simple model of the expansion rate, we assume that the distribution of the axial magnetic field energy density has a Gaussian profile, }
\begin{eqnarray}\label{Eq:uB}
     u_B(x,y,z,t) &=& u_0\frac{V_0}{ \sigma_x \sigma_r^2 }\exp \left[ -\frac{(x - x_0)^2}{2\sigma_x^2} - \frac{r^2}{2\sigma_{r}^2}  \right] \\
     \sigma_{x}(t) &=& \sigma_{x_0}+\nu_x t,~ ~\sigma_{r}(t)~=~\sigma_{r_0}+\nu_r t ,
\end{eqnarray}
\rc{with widths along different directions, $\sigma_x(t)$ and $\sigma_r(t)$, that vary over time to represent the expansion effects. 
The normalized factor $V_0/(\sigma_x \sigma_r^2 )$ ensures that the total energy associated with the magnetic field, $U_B = \int u_B(x,y,z,t)dV=\int B_{x}^{2}/(2 \mu_0)dV$, remains constant during the expansion. The peak of the axial magnetic field is assumed to be at $x = x_0$, $r = 0$. We determined the parameters in Eq.~(\ref{Eq:uB}) through several steps. First, we used simulation data at \(t = 20\) fs to calculate the energy of the axial magnetic field as a function of \(x\) and \(r\), assuming axial symmetry. From the resulting profile, we found \(\sigma_{x}(t = 20~{\rm fs}) = 3.5~[\micron]\), \(\sigma_{r}(t = 20~{\rm fs}) = 4.5~[\micron]\), and \(u_0V_0 = 4.4~[\rm mJ]\). We then scanned the parameters \(\nu_x\) and \(\nu_r\) to achieve the best fit to the blue line in \cref{ebEth}(a), with the fitting result shown as a black dashed line with cross marks. The fitting parameters were determined to be \(\nu_x = 2.7\times10^{-2}~[\micron\cdot\rm fs^{-1}] \sim 0.09c\) and \(\nu_r = 4.5\times10^{-3}~[\micron\cdot\rm fs^{-1}] \sim 0.015c\). Consequently, we obtained \(\sigma_{x}~{[\micron]}= 3.0  + 0.027t~{[\rm fs]}\) and \(\sigma_{r}~{[\micron]} = 4.4 + 0.0045t~{[\rm fs]}\). Our model shows that the expansion speeds differ significantly between the longitudinal and transverse directions, which can be partially attributed to the confinement by the generated axial magnetic fields. }

\rc{In the proposed expansion model, the axial magnetic field, whose energy density is described by Eq.~(\ref{Eq:uB}), is given by }
\begin{equation}\label{Eq:avB}
     B(x,y,z,t) = \sqrt{2 \mu_0 \frac{u_0V_0}{ \sigma_x \sigma_r^2} } \exp\left [-\frac{(x - x_0)^2}{4\sigma_x^2} - \frac{r^2}{4\sigma_{r}^2}  \right].
\end{equation}
\rc{After substituting all the parameters, we obtain the following explicit form:}
\begin{eqnarray}
     B_x(x,y,z,t)~{[\rm kT]} = \frac{105}{(4.4+0.0045t~[{\rm fs}])\sqrt{3.0+0.027t~[{\rm fs}]}} \times\nonumber\\
     \exp\left[-\left(\frac{(x - x_0)~[\micron]}{6.0+0.054t~[{\rm fs}]} \right)^2- \left(\frac{r~[\micron]}{8.8+0.009t~[{\rm fs}]} \right)^2 \right].
\end{eqnarray}
\rc{For the peak of the axial magnetic field $B_x^{m}$, the  decay scaling obtained from our model is }
\begin{equation}\label{Eq:Bscaling}
B_x^{m}~[{\rm kT}] = 105(4.4+0.0045t~[{\rm fs}])^{-1} (3.0+0.027t~[{\rm fs}])^{-0.5}
\end{equation}
\rc{The result of \cref{Eq:Bscaling} is represented by the black dashed line with cross marks in \cref{ebEth}(b).  The fitting model results  in \cref{ebEth}(a) and \cref{ebEth}(b) are shown for $t$ ranging from $t$ = 0~fs to $t$ = 1700~fs. Since the simulation data ends at $t = 920$~fs, our model provides a longer prediction of the magnetic decay. }

\bigskip
\subsection{Azimuthal current distribution}
\begin{figure*}[htb]
    \centering
    \includegraphics[width=0.8\textwidth]{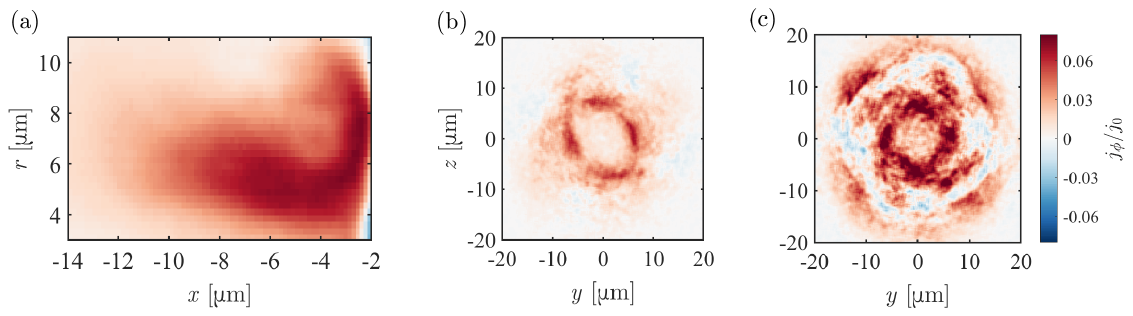}
    \caption{Azimuthal current density $j_{\phi}$ (normalized to $j_0 = -|e|c n_c = - 4.8\times 10^{16}~\rm A\cdot m^{-2}$) at $t = 20$~fs. (a) Azimuthal current density $j_{\phi}$ averaged over $\phi$ as a function of $x$ and $r$. (b) and (c) Two-dimensional transverse slices of $j_{\phi}$ at $x = -10$ and $-5~\micron$, respectively.}
    \label{j_r2d}
\end{figure*}

We begin by analyzing the azimuthal current density $j_{\phi}$, which is believed to be responsible for the generation of the axial magnetic field. 
\Cref{j_r2d} (a) illustrates the distribution of the azimuthal current density $j_\phi$ (averaged over the azimuthal angle) for $\varphi = 0.28 \pi$ at $t = 20$~fs as a function of axial $x$ and radial $r$ positions.
The current density is normalized to $j_0 = -|e|c n_c  = - 4.8\times 10^{16}~\rm A\cdot m^{-2}$ which corresponds to the electrons with density $n_e = n_{c} = 1.0\times 10^{21}$~cm$^{-3}$ rotating at the speed of light. 
The establishment of the strong azimuthal current $j_\phi$ shown in \cref{j_r2d} (a) is crucial for the efficient generation of the strong axial magnetic field in our multi-beam approach. \rc{The negative azimuthal current, $j_{\phi} < 0$ or $j_{\phi}/j_0 > 0$, should produce a negative axial magnetic field [note that $j_0 < 0$].  }
This is the same direction as shown in \cref{bx3d&r} (a).
Two-dimensional transverse slices of $j_\phi$ at $x = -10$ and $-5~\micron$ are presented in \cref{j_r2d} (b) and (c), respectively. 
As observed in \cref{j_r2d} (a), the azimuthal current density $j_\phi$ at $x=-5~\micron$ is stronger compared to the current density at $x=-10~\micron$.

To estimate the maximum value of $|B_x|$, we make an order of magnitude assumption that $j_\phi$ is uniform within a cylindrical region of radius $R$ and length $2h$. 
After applying the Biot-Savart law~\cite{Jackson1975}, we obtain
\begin{eqnarray}\label{eq:bx_cal}
    \max|B_x|&\approx&\dfrac{\mu_0}{2}\int_0^R\int_{-h}^{h}|j_{\phi}|\dfrac{r^2}{(r^2 + x^2)^{3/2}}\mathrm{d}x\mathrm{d}r \nonumber \\
    &= &\mu_0|j_{\phi}|h \operatorname{arsinh}(R/h),
\end{eqnarray}
where $\mu_0 = 1.26 \times 10^{-6}$~H/m is the permeability of free space. 
From the information provided in \cref{j_r2d} (a), we can set $R \approx h \approx 5~\micron$. 
Considering that the value of the azimuthal current density in \cref{j_r2d} (c) is $|j_{\phi}| \approx 0.08~|j_0|$, where $j_0 \equiv -|e| c n_c$, we can estimate the maximum value of $|B_x|$ as $\max |B_x| \simeq 20$~kT. This result is close to the peak magnetic field strength $ \left<B_{x}\right> \simeq 1.8~B_0 = 18$~kT observed in \cref{bxbar_t} (b). 

Assuming that the motion of the electrons is the main contributor to the current, we can calculate the effective azimuthal velocity as $v_{\phi} \approx - j_{\phi}/|e|n_e$. Furthermore, the azimuthal current density can be used to estimate the OAM density of hot electrons.
With the electron density obtained from the simulations, $n_e \approx 10^{27}$ m$^{-3}$, we find that the rotation velocity is about $v_{\phi} \approx 0.1c$, which implies a fast rotating plasma environment. {One may wonder if the rotating effect is just a combination of four cross currents forming a shape similar to a square. We can estimate the expanding effect of the cross currents (if these currents are indeed present) using $v_{\phi} \approx 0.1c$. After 400~fs (from $t=20$~fs to $t=420$~fs), the transverse shift should be $R_S = 0.1c \times 400~{\rm fs} \approx 12~\micron$. Such a significant shift is not observed in our simulation, which leads us to conclude that we are indeed dealing with a fast rotating plasma rather than with four cross currents.}
We can write the OAM density of electrons as $L_{xe} \approx r \gamma_{a} m_e n_e v_{\phi}$, where $\gamma_{a}$ is the relativistic gamma-factor and $n_e$ is the electron density. We next take into account that $j_{\phi} \approx |e| n_e v_{\phi}$ to obtain that $L_{xe} \approx r \gamma_{a} m_e j_{\phi}/|e|$. With $r = f_0$ and $\gamma_{a} \approx (1 + a_0^2)^{1/2} \approx 8$, the OAM density is approximately $L_{xe} \approx 2.4~\kgms$. 
Compared to the case where twisted lasers are employed to generate OAM~\cite{Shi2018}, our scheme produces a rotating plasma environment with electron density and rotation velocity that are two orders of magnitude higher.

\subsection{OAM distributions analysis}

\begin{figure*}[htb]
    \centering
    \includegraphics[width=0.8\textwidth]{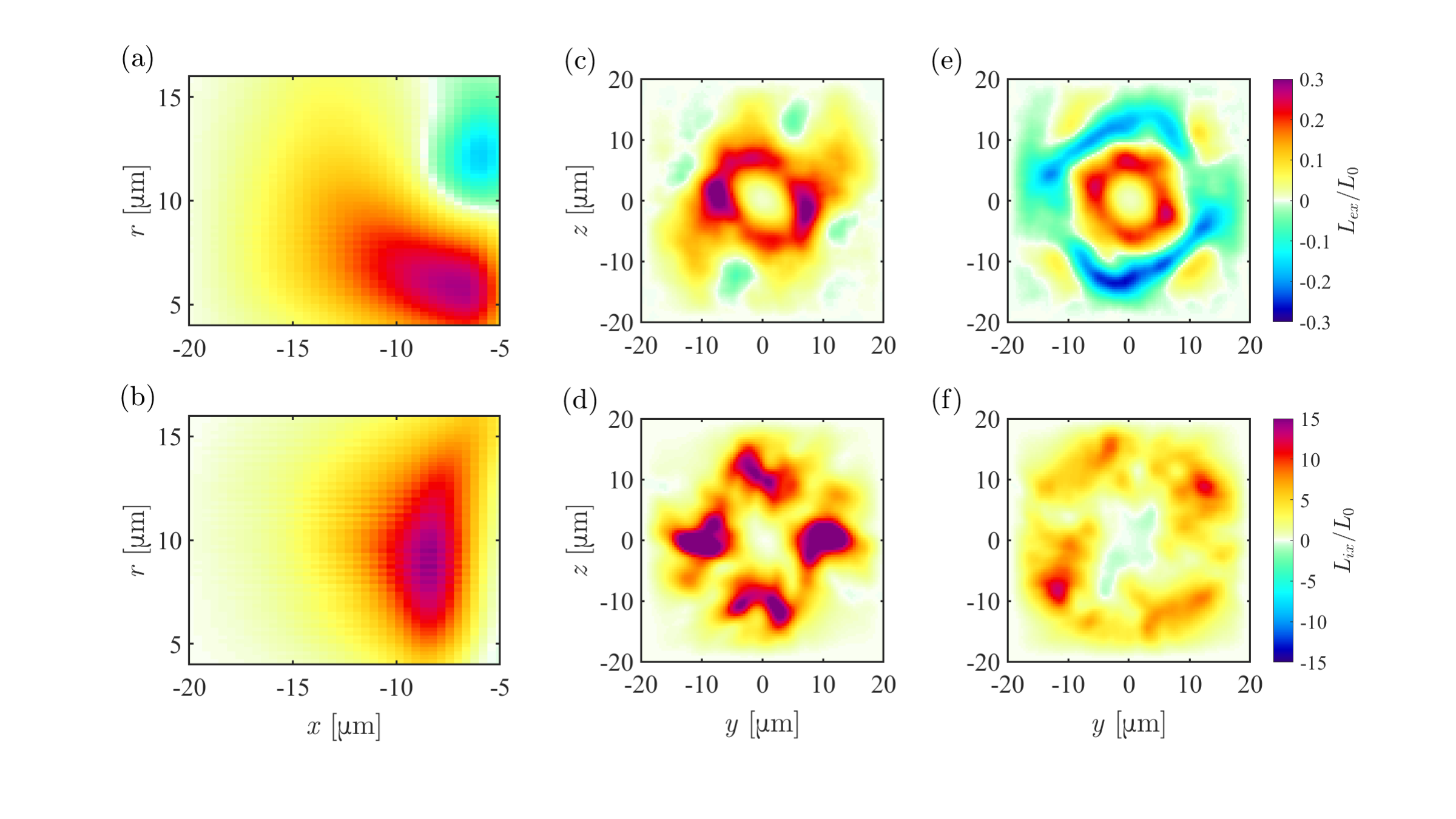}
    \caption{Density of axial OAM for electrons (top row) and ions (bottom row) normalized to $L_0 \equiv n_c m_e c w_0 \approx 1.65~\kgms$ at $t = 20$~fs in the simulation with the twist angle $\varphi = 0.28\pi$. (a), (b) Azimuthally averaged densities of axial OAM $L_{xe}$ and $L_{xi}$ as functions of $x$ and $r$. (c), (d) and (e), (f) Transverse slices of $L_{xe}$ and $L_{xi}$ at $x = -10$ and $-5~\micron$, respectively.}
    \label{oamxei}
\end{figure*}

To study the rotational effect induced in the plasma and to quantify the AM transfer from the four laser beams with a twist angle of $\varphi = 0.28\pi$ in our simulation, we analyze the density of axial OAM in electrons ($L_{xe}$) and ions ($L_{xi}$). Both quantities are normalized to a reference value $L_0 \equiv n_c m_e c w_0 \approx 1.65~\kgms$, where $L_{0}$ represents electrons with density $n_c$ rotating with azimuthal velocity $v_{\phi} = c$ at a radial position with $r = w_0$.
The analysis is conducted at a time of $t = 20$~fs. 
In \cref{oamxei}, the densities of axial OAM for electrons (top row) and ions (bottom row) are shown at $t = 20$~fs. \Cref{oamxei} (a) and (b) show angle-averaged $L_{xe}$ and $L_{xi}$ as functions of $x$ and $r$. Further insight can be provided by two-dimensional transverse slices of $L_{xe}$ ($L_{xi}$) at two different axial locations, $x = -10~\micron$ and $x = -5~\micron$, shown in \cref{oamxei} (c) and \cref{oamxei} (e) (\cref{oamxei} (d) and \cref{oamxei} (f)), respectively.  
{We would like to point out that the OAM density in Fig.~C1 from the appendix of our previous work~\cite{Shi2023} is artificially high due to a data processing error.}

\rc{As can be seen from \cref{oamxei} (a) and (b)}, the axial OAM density distribution range of ions in the longitudinal direction is smaller than that of electrons.
This behavior could stem from the challenges involved in altering the motion state of ions. While the axial OAM density of electrons remains positive and gradually decreases away from the target in most of the region ($x < -7.5~\micron$), another intriguing phenomenon is the occurrence of negative axial OAM density for electrons near the preplasma ($x\in[-7.5,-5]~\micron$), indicating a reversal of electron rotation at larger radial positions.  
 Over time, the hot electrons undergo radial and longitudinal expansion away from the target, which is likely to contribute to the reduction in axial magnetic field strength. According to \cref{oamxei},  we can approximate the distribution ratio of the OAM density between electrons and ions as $\eta_{ei} = L_{xe}/L_{xi} \approx 1\%$. We attribute this observation to the significant mass disparity between ions and electrons (where $m_i/m_e\approx 10^4$). Despite the higher OAM density of the ions, the velocity of the ions is still much less than that of the electrons. Since the current does not depend on particle masses, these simulation results confirm our earlier assumption that the dominant contribution to the azimuthal current component comes from electron motion.

 To construct an analytical model for this new scheme, we begin by examining the AM density of the four laser pulses. This AM density, denoted as $L_x = \varepsilon_0(\bm r\times[\bm E\times \bm B])_x$, vanishes on the axis and reaches its maximum value at the beam offset. In light of the extensive simulation results we presented earlier, we now turn our attention to the process of AM absorption. 
 We envision this absorption occurring within a cylindrical region of radius $R$ and \rc{length $2\Delta h$}.
 The lasers, first and foremost, impart AM to electrons. Through the electron current, a strong axial magnetic field is generated. During the generation phase, the evolving magnetic field introduces an azimuthal electric field $E_\phi$. This electric field, in turn, facilitates the transfer of absorbed OAM from electrons to ions. {The electrons and ions are assumed to rotate rigidly in a shell with angular velocities $\zeta_e$ and $\zeta_i$, respectively. Since the particles are moving away from the center, we expect it to work well only for estimating initial peak values.}
 We introduce OAM to describe the rotational motion of electrons and ions, denoted as $L_e = I_e\zeta_e$ and $L_i = I_i\zeta_i$ respectively. 
 Here, $I_e$ is the moment of inertia for electrons and is given by \rc{ $I_e = \pi R^4 \Delta h m_e n_e$ }.
 Meanwhile, $I_i$ represents the moment of inertia for ions and is calculated as $I_i = (m_i/Zm_e)I_e$, where $Z$ is nuclear charge number of ions. These OAM estimates capture the rotational dynamics of the charged particles. 
 The overall evolution of the OAM of electrons and ions can be described by the following equations~\cite{Liseykina2016,Popruzhenko2019}
\begin{eqnarray}\label{amevl}
    I_e\frac{d\zeta_e}{dt}=M_{\rm abs}-M_{\rm E},~~I_i\frac{d\zeta_i}{dt}=M_{\rm E},
\end{eqnarray}
where $M_{\rm abs}$ represents the torque resulting from OAM absorption. 
The term $M_{\rm E}$ signifies the torque due to the azimuthal electric field $E_\phi$ and can be expressed as $M_{\rm E}=\int |e|E_\phi(r)rn_ed^3r \approx |e|E_\phi(R)I_e/(m_eR)$.
The rotational motion of electrons generates current density $j_{\phi e}$, which can be approximated as $j_{\phi e}\simeq-|e|n_e\zeta_eR$. 
Furthermore, we can estimate the magnitude of $E_{\phi}(R) \approx -R (\partial B_x/ \partial t) $ by using the Eq.~(\ref{eq:bx_cal}).
As a result, we can express $M_{\rm E}$ as $M_{\rm E}\simeq (d\zeta_e/dt)I_e'$, where \rc{ $I_e'\equiv \omega_p^2 R \Delta  h \operatorname{arsinh}(R/\Delta h)I_e/c^2$ }, $\omega_p = (n_e e^2/m_e\varepsilon_0)^{1/2}$ represents the plasma frequency.
According to the parameters above, we can get the result of $I_e^{'} \gg I_e$.  
With these considerations in mind, we arrive at the expression $\zeta_e(t)=(I_e+I_e')^{-1}\int_0^tM_{\rm abs}(t')dt'$.
This equation illustrates that the rotation of electrons closely follows the temporal profile of $M_{\rm abs}(t)$, akin to an effect of effective inertia. Given that $I_e^{'} \gg I_e$ under the conditions of our study, we find that $M_{\rm E}$ is approximately equal to $M_{\rm abs}$. Consequently, we can get the result of $L_i\simeq\int_0^tM_{\rm abs}(t')dt'\simeq L_e I_e'/I_e \gg L_e$.
This result reveals that the total OAM of ions greatly exceeds that of electrons, consistent with the simulation findings presented in \cref{oamxei}.

The absorption of electromagnetic AM is proportional to energy absorption in general. Like other studies on IFE, the transfer of AM from the four laser beams to the electrons can be assessed through the application of AM conservation principles~\cite{Haines2001, Ali2010, Longman2021, Shi2023}. 
{The number of absorbed photons is  $N_{\rm abs} = U_{\rm abs}/\omega$, where $U_{\rm abs}$ represents the absorbed energy, approximately $U_{\rm abs} \simeq \eta_{\rm abs}U_{\rm L}$.} 
Here, $\eta_{\rm abs}$ is the absorption coefficient of laser intensity absorbed over the axial distance, assumed to conform to $\eta_{\rm abs} = \eta_{0} \exp(x/ \mathcal{\kappa})$. 
The absorbed OAM carried by laser photons is subsequently transferred to both electrons and ions, with the fraction of the OAM carried by electrons denoted as $\eta_{ei} \equiv L_{xe}/L_{xi}$. Since most of absorbed OAM is transferred to ions, we can get the OAM density of electrons as $L_{xe} \simeq \eta_{ei} L_{\rm abs}$.
Specifically, the axial OAM density of electrons can be approximated as follows:
\begin{eqnarray}\label{oam_abs}
    L_{xe}(x, r) &\approx& 0.75\eta \eta_{ei}\frac{I_0\tau_g}{\kappa c} \sin(\theta) \sin(\varphi) \mathscr{F}_{xr}, \nonumber \\
    \mathscr{F}_{xr} &=& r \exp({x/\kappa- 2(r - f_0)^2/w_0^2}),
\end{eqnarray}
where $I_0$ represents the peak intensity of the incident laser pulses, and $\tau_g$ stands for their duration. 
For the sake of simplicity, we deduce from the simulation results that $\eta = \int_{x_e}^{0} \eta_{\rm abs} dx \simeq \kappa \eta_0 \approx 0.1$ ($\kappa \approx 3 ~\micron$). 
We utilize $r = 6~\micron$ and $x = 0~\micron$ to determine the peak OAM density of the electrons, $L_{xe} \approx 1.75~\kgms$. 
This outcome falls within a similar range as the peak OAM density ($L_{xe}\approx 1.69~\kgms$) presented in \cref{oamxei}. Furthermore, it closely aligns with the result ($L_{xe}\approx 1.5~\kgms$) obtained through the utilization of azimuthal current density $j_{\phi}$ as demonstrated in \cref{j_r2d} considering the roughness of the model. 

To simplify the analysis, we have made some approximations for the absorption coefficient $\eta_{\rm abs}$, assuming its independence from the laser peak intensity $I_0$ and incidence angles (the polar angle $\theta$ and azimuth angle $\varphi$). In reality, the absorption mechanism is more intricate~\cite{Brunel1988,Levy2014,Grassi2017}. 
\rc{By utilizing the OAM absorbed by electrons $L_{xe}$ and the associated azimuthal current density, we thus estimate the final axial magnetic field as
\begin{eqnarray} \label{bx_theory}
    \left|\frac{B_{x}}{B_0}\right|\propto \left|\frac{j_{\phi}}{j_0}\right|\propto \eta \eta_{ei}\frac{a_0^2}{\gamma_{a}}\frac{c\tau_g}{\kappa} \frac{\mathscr{F}_{xr}}{r} \sin(\theta) \sin(\varphi). 
\end{eqnarray}
}
In accordance with Eq. \eqref{bx_theory}, the manipulation of the axial magnetic field is achievable by altering the sign of the twist angle $\varphi$, a proposition validated in the subsequent section.

\section{\label{Sec-4} Impact of laser parameters and target geometry on the axial magnetic field generation}

The purpose of this section is to examine various aspects of our scheme that are likely to be important during experimental implementation at multi-kJ PW-class laser facilities such as the upcoming upgrade of SG-\Rmnum{2}. The comprehensive analysis presented in this section is based on a series of 3D PIC simulations and it addresses such factors as twist angle, polarization direction, phase, delay, and plasma front structure.
We want to point out that we use a target with a preplasma and laser beams with wavelength of $1.053~\micron$, which differs from the use of nanowires and laser wavelength of $0.8~\micron$ in our previous work~\cite{Shi2023}.

\subsection{Twist angle dependency} \label{Sec-4A}

To validate the significance of the twist angle $\varphi$, we conducted two additional 3D PIC simulations. These simulations encompassed twist angles of $\varphi = 0$ and $\varphi = -0.28 \pi$, the latter being opposite in twist direction to the original setup. These additional simulations were performed while keeping all other parameters the same as listed in \cref{LPI par}.
\begin{figure*}[htb]
    \centering
    \includegraphics[width=0.8\textwidth]{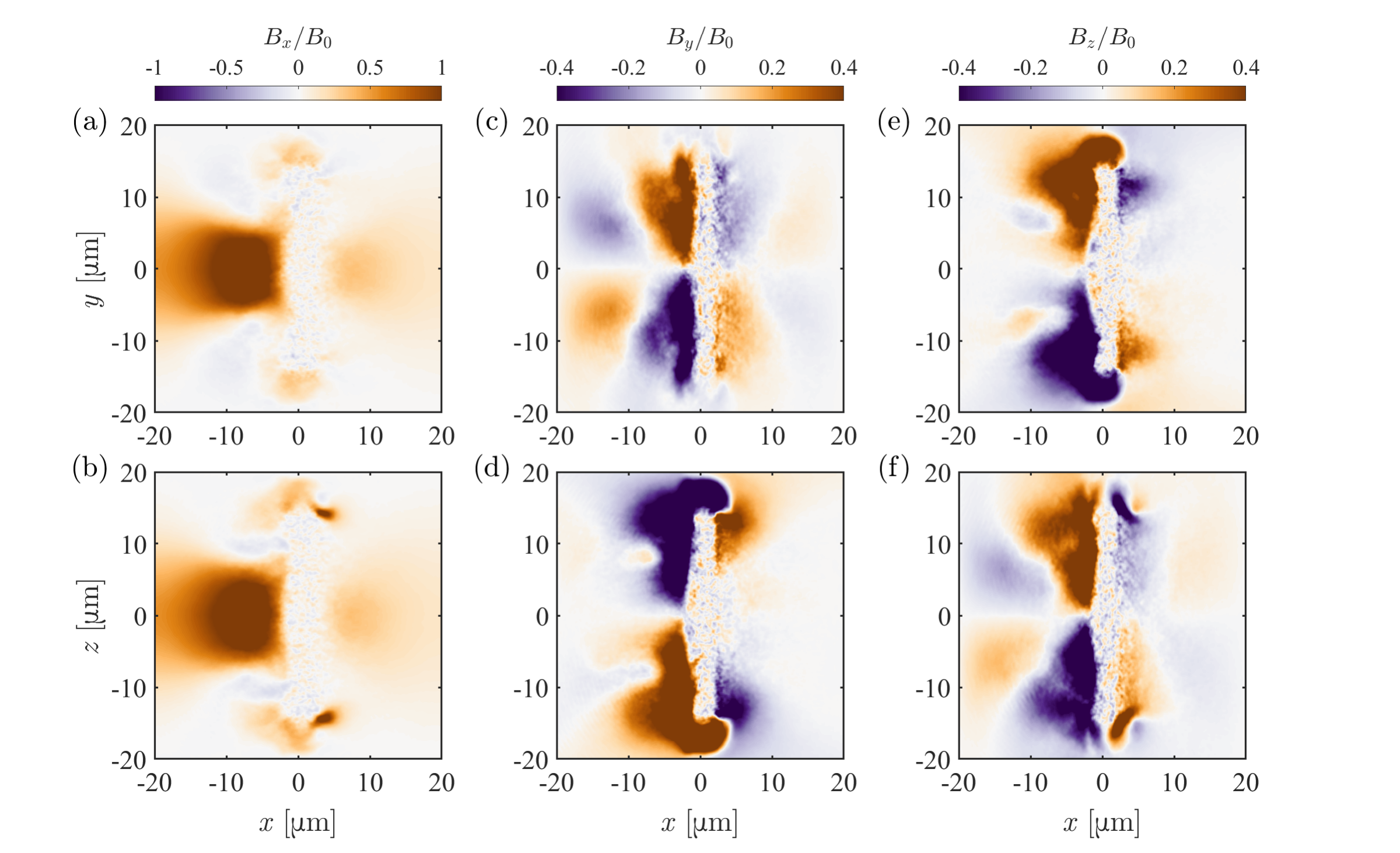}
    \caption{Simulation result for laser beams \textit{with a reversed twist} ($\varphi = -0.28\pi$). Two-dimensional longitudinal slices at $t = 20$~fs of all three components of the time-averaged (averaged over 20 fs output dumps) magnetic field in the 3D PIC simulation with the parameters listed in \cref{LPI par}.
    The fields are normalized to $B_0=2\pi m_ec/|e|\lambda=10.0~\rm kT$. $B_x$, $B_y$, and $B_z$ are shown in $(x,y)$-plane (top row) and $(x,z)$-plane (bottom row), respectively. Note that the color scale for $B_x$ is different from that for $B_y$ and $B_z$.}
    \label{bxyz2d_re}
\end{figure*}

\begin{figure*}[htb]
    \centering
    \includegraphics[width=0.8\textwidth]{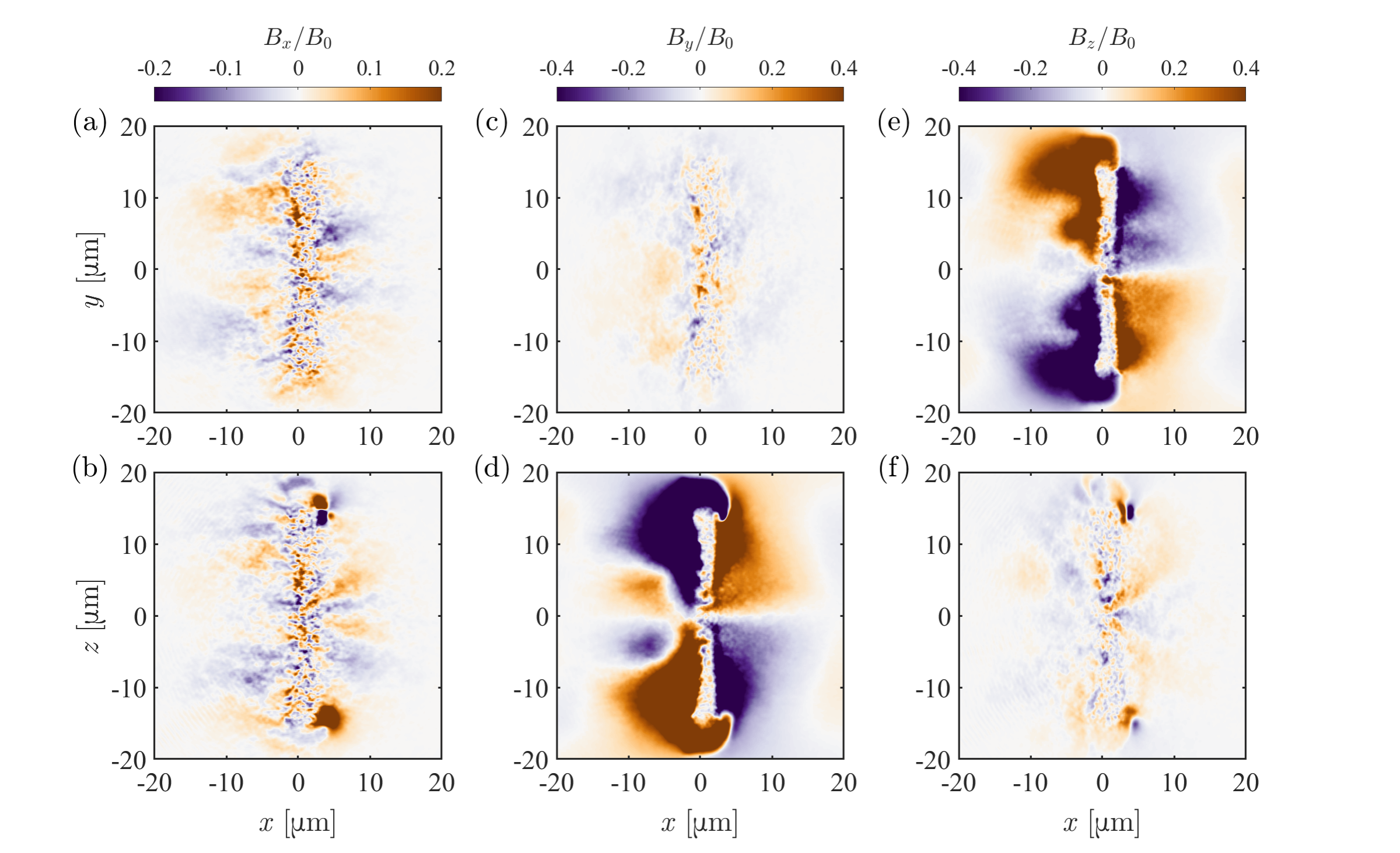}
    \caption{Simulation result for laser beams \textit{without a twist} ($\varphi = 0.0\pi$). Two-dimensional longitudinal slices at $t = 20$~fs of all three components of the time-averaged (averaged over 20 fs output dumps) magnetic field in the 3D PIC simulation with the parameters listed in \cref{LPI par}. The fields are normalized to $B_0=2\pi m_ec/|e|\lambda=10.0~\rm kT$.  $B_x$, $B_y$, and $B_z$ are shown in $(x,y)$-plane (top row) and $(x,z)$-plane (bottom row), respectively. Note that the color scale for $B_x$ is different from that for $B_y$ and $B_z$.}
    \label{bxyz2d_no}
\end{figure*}

\Cref{bxyz2d_re} and \Cref{bxyz2d_no} present the two-dimensional longitudinal slices of all three components of the time-averaged (averaged over 20 fs output dumps) magnetic field at $t = 20$~fs with the twist angle $\varphi = -0.28\pi$ and $\varphi = 0.0\pi$, respectively. When employing laser beams with the opposite twist, the axial magnetic field is reversed, as depicted in \cref{bxyz2d_re}. 
Comparing \cref{bxyz2d_re} and \cref{bxyz2d}, it is evident that the axially symmetric axial magnetic field $B_x$ is reversed, while the transverse magnetic fields $B_y$ and $B_z$ still exhibit anti-symmetric features. 
Moreover, the directions of $B_y$ in the $(x,y)$-plane and the directions of $B_z$ in the $(x,z)$-plane are reverse in \cref{bxyz2d_re} compared to \cref{bxyz2d}. In \cref{bxyz2d_no}, the results suggest that in the absence of a twist, i.e., $\varphi = 0$, the axial magnetic field diminishes substantially, while the effect on the transverse magnetic fields is relatively less pronounced due to the existence of a longitudinal current even in the absence of a twist. 

The generation of transverse magnetic fields occurs in all three scenarios due to the presence of an axial current induced by the laser pulses. These transverse fields are particularly weak at small radii, underscoring the dominance of the axial magnetic field in governing the magnetic field within the central region. This observation confirms the recognition of a robust mechanism responsible for the generation of axial magnetic fields. It also suggests that the twist parameter can act as a knob for controlling the magnetic field within the central region, similar to the role of the topological charge $l$ for LG beams in the IFE.

\subsection{Polarization direction dependency}

The polarization direction may play a role in the AM transfer process of our scheme that employs multiple LP Gaussian beams.  Intuitively, it may seem that the direction of the transverse laser electric field is crucial for the generation of the azimuthal current. However, it is worth noting that the laser electric field is oscillating, while the current generated by this scheme is quasi-static. The quasi-static current and magnetic field evolve on a much longer time scale than the laser period. The electric field oscillations make it difficult for the laser beams with specific polarization directions to generate the quasi-static azimuthal current with a preferred direction. However, the polarization direction can affect the absorption of laser energy and AM. Based on the absorption characteristics of the plasma, different polarization directions result in different degrees of laser energy absorption within the plasma \cite{Meyerhofer1993,Pearlman1977,JS.Pearlman1977,Balmer1977}. This variation could potentially affect the distribution of hot electrons generated within the plasma.

\begin{figure*}[htb]
    \centering
    \includegraphics[width=0.8\textwidth]{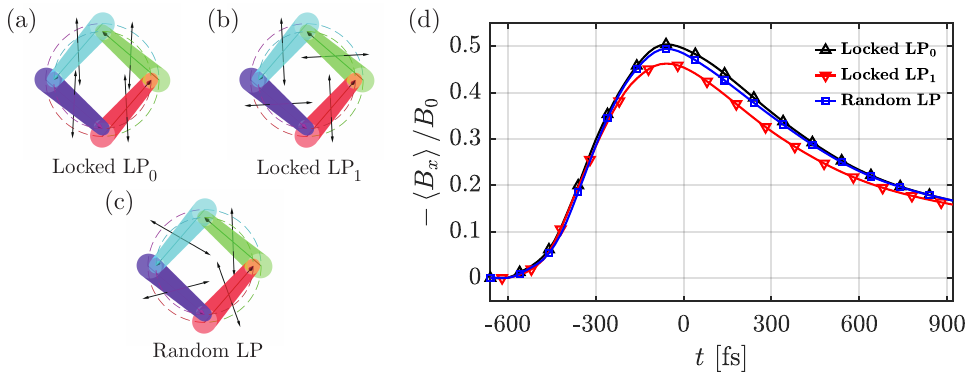}
    \caption{Two-dimensional view of the spatial arrangement and polarization direction of the four laser beams in the $(y,z)$-plane. (a) Locked~LP$_0$ simulation: The linear polarization directions of all four laser beams are along the $y$-axis. (b) Locked~LP$_1$ simulation: A pair of laser beams (cyan and red) are polarized along the $y$-axis, whereas the other pair (green and purple) are polarized along the $z$-axis. (c) Random LP simulation: Random phase shifts $\xi \in (0, \pi)$ are added to the initial polarization directions of the four laser beams, disrupting the regular polarization arrangement of the lasers. (d) Average axial magnetic field within a rectangular volume of length $a=17.9~\micron$ ($x\in[-19.9,-2.0]~\micron$), width and height $b$, $c=16~\micron$ ($y$, $z\in[-8.0,8.0]~\micron$) as a function of time, where $t=0~\rm fs$ is defined as the moment when the lasers leave the simulation space completely. The black, red, and blue solid lines represent the cases of Locked~LP$_0$, Locked~LP$_1$, and Random~LP, respectively.}
    \label{LP_d}
\end{figure*}

To investigate the impact of the direction of laser polarization, we performed two additional 3D PIC simulations where the polarization direction was deliberately changed compared to that used in the original simulation presented in \cref{Sec-3}. \Cref{LP_d} shows the direction of the transverse electric field in the $(y,z)$-plane for each beam, where \cref{LP_d} (a) shows the original polarization and \cref{LP_d} (b) and \cref{LP_d} (c) show the polarization in the two additional simulations. In the original simulation discussed in \cref{Sec-3}, we set the polarization of all four laser beams such that the transverse electric field of each beam in the $(y,z)$-plane was directed along the $y$-axis, as shown in \cref{LP_d} (a). We will refer to this simulation as ``Locked LP$_0$''. 
In the first additional simulation, we rotate the polarization direction in every other beam by $\pi/2$. The resulting polarization is shown in \cref{LP_d} (b). We refer to this simulation as ``Locked LP$_1$''. In the second additional simulation, we introduced random rotation $\xi \in (0, \pi)$ to the original polarization direction for each laser beam. The polarization used in the simulation is shown in \cref{LP_d} (c). We refer to this simulation as ``Random LP''. The purpose of this simulation is to examine whether the multi-beam scheme remains effective even when the linear polarization directions are randomly disordered.  It is important to note that in this article we are primarily focused on the effects of the combination of four laser beams with linear polarization on the generation of axial magnetic fields. We do not consider other polarization states. This choice is mainly due to the design of the multi-beam LP laser setup used in most high-power laser systems.

\Cref{LP_d} (d) shows the time evolution of the spatially averaged magnetic field for the three different polarization setups:  Locked LP$_0$ (black), Locked LP$_1$ (red), and Random LP (blue).
It is evident that the blue and black curves almost entirely overlap, whereas the red curve is slightly lower than the other two. 
Based on these results, we infer that co-polarized beams (beams with the same polarization direction) are slightly more favorable for generating the axial magnetic field in our scheme. The result also suggests that even if the polarization directions are somewhat disordered, it will not significantly impact the effectiveness of our scheme. \rc{From the perspective of AM absorption, the axial AM carried by the four laser beams, as given by \cref{eq: 1}, depends solely on the longitudinal component of the wave vector. Therefore, as long as the twisted pointing directions of the four lasers remain unchanged, polarization changes should have a minimal effect on AM transfer and, consequently, on the magnetic field generation mechanism. } This result is promising for experimental implementation, potentially eliminating the need for deliberate adjustments of initial linear polarization directions.


\subsection{Delay dependency}

\begin{figure*}[htb]
    \centering
    \includegraphics[width=0.8\textwidth]{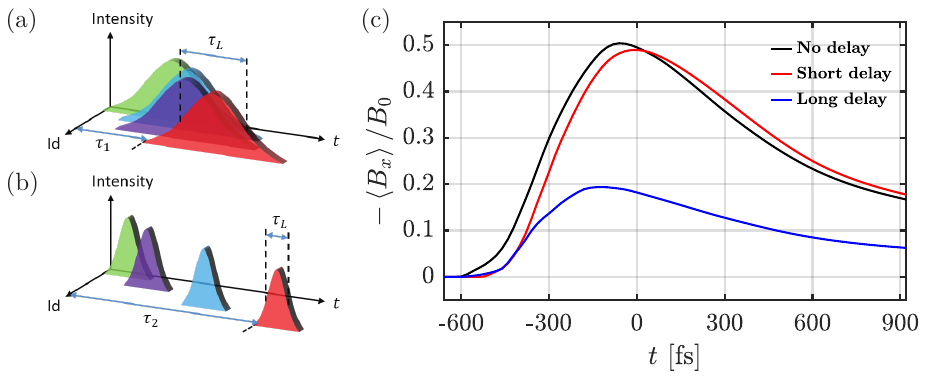}
    \caption{(a) and (b) Two delay scenarios considered in the simulations: the short delay and the long delay. The individual pulse duration is $\tau_L = 600~\rm{fs}$ and the considered delays are 
    $\tau_1 = 250~\rm{fs}$ and $\tau_2 = 1~\rm{ps}$. Note that the delays are random in the short delay scenario, whereas they are fixed in the long delay scenario. The delays are shown on different time scales. (c) Average axial magnetic field within a rectangular volume of length $a=17.9~\micron$ ($x\in[-19.9,-2.0]~\micron$) and width and height $b$, $c=16~\micron$ ($y$, $z\in[-8.0,8.0]~\micron$) as a function of time, where $t=0~\rm fs$ is defined as the moment when the lasers leave the simulation space completely. The black, red, and blue solid lines represent the cases with no delay, a short delay, and a long delay, respectively.}
    \label{D_contr}
\end{figure*}

In high-power laser experiments, time delay can pose several challenges that affect the accuracy and reproducibility of experimental outcomes. Specifically, a time delay can cause the interactions between the individual laser beams and the target to occur at different times. This may potentially impact the interaction dynamics. Given the challenges posed by laser synchronization difficulties, it is valuable to examine the role of delays in our scheme. 
To address this, we examined two different delay scenarios in simulations. In the first scenario, we introduced random delays of up to 250 fs for each of the four laser beams, representing the short delay case. 
In the second scenario, the delays of four beams are set to 0~fs, 150~fs, 600~fs, and 1000~fs.
The delay settings for the two scenarios are illustrated in \cref{D_contr} (a) and \cref{D_contr} (b), where the duration of each laser pulse is 600~fs.

We find that under the short delay condition in the original simulation, the interaction between the four laser beams and the plasma leads to generation of hot electrons over a greater spatial extent, implying a higher energy transfer efficiency and much faster expansion of the hot electron population. 
Consequently, in the context of the short delay scenario, various factors, including the strength and spatial range of the axial magnetic field, exhibit a magnitude that is one order higher than those observed under the long delay scenario. 

\Cref{D_contr} (c) provides the temporal evolution of the spatially averaged axial magnetic field for the two considered delay scenarios. The field for the case without the delay, i.e. the ``no delay'' curve, is given for reference.  
We find that the differences between the short delay case and the no delay case are minimal. However, the long delay noticeably reduces the strength of the generated magnetic field compared to the case without the delay. The peak value reaches 40\% of the peak value without the delay. Despite this reduction, the magnetic field still reaches the kilotesla range, demonstrating that our scheme remains effective under sufficiently long delay conditions. This provides encouraging prospects for experimental implementation.

\rc{An important takeaway from the 'long delay' simulation is that a strong magnetic field can be generated even if not all four beams temporally overlap. In our case, only two beams overlap, as seen in \cref{D_contr}(b). However, as shown by \cref{eq: 1} and the corresponding discussion in \cref{Sec-2}, two lasers with opposite orientations are sufficient to provide axial angular momentum. In our simulation, there is a significant period during which at least two lasers interact with the plasma target simultaneously, ensuring the transfer of angular momentum and thereby supporting magnetic field generation.}


\subsection{Phase dependency}

\begin{figure*}[htb]
    \centering
    \includegraphics[width=0.9\textwidth]{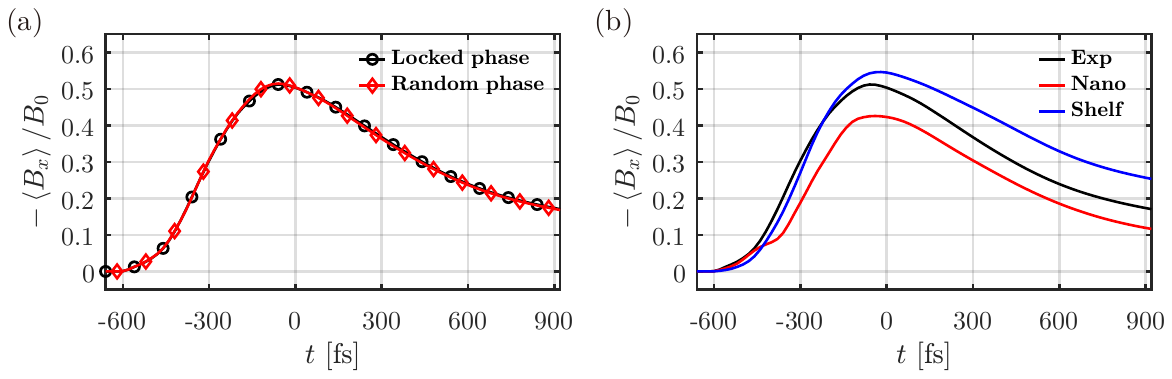}
    \caption{Average axial magnetic field within a cuboid of length $a=17.9~\micron$ ($x\in[-19.9,-2.0]~\micron$), width and height $b$, $c=16~\micron$ ($y$, $z\in[-8.0,8.0]~\micron$) as a function of time, where $t=0~\rm fs$ is defined as the moment when the lasers leave the simulation space completely. (a) The locked phase case is depicted by the solid black line with round markers, while the random phase case is shown as the solid red line with diamond markers. \rc{(b) Impact of the density profile at the laser-irradiated surface. The shelf density profile, the exponential density profile, and the nanowire target are represented by the blue line, the black line, and the red solid line, respectively}.}
    \label{Phs_contr}
\end{figure*}

In multi-beam laser experiments, the phase relationship between different laser beams can produce different synthetic effects. For example, multiple laser beams with a locked phase difference can produce an interference pattern and then improve the efficiency of laser energy conversion into hot electrons~\cite{Morace2019}. To investigate the effect of phase locking between four beams in our scheme, we introduce random initial phases $\psi_{rd}\in(0,2\pi)$ for each laser beam. \Cref{Phs_contr} (a) shows the time evolution of the spatially averaged magnetic field for two different cases: the locked phase and the random phase. We find that the introduction of random phases does not cause any significant changes. We can conclude that the phase relationship in the simulation did not have a significant effect on the generation of the axial magnetic field.
No phase control is required for the combination of multiple
laser pulses in our scheme.

\rc{\subsection{Impact of pointing stability}}

\rc{Laser pointing fluctuations are inherent in any laser system, so it is important to understand their impact. These fluctuations can affect our scheme by altering the beam offset ($f_0$), the azimuthal angle ($\varphi$), and the beam convergence angle ($\theta$) for each beam. Our approach is to treat fluctuations in the beam offset as independent from those in $\varphi$ and $\theta$, allowing us to assess their relative importance.}

\begin{figure}[htb]
    \centering
    \includegraphics[width=0.9\columnwidth]{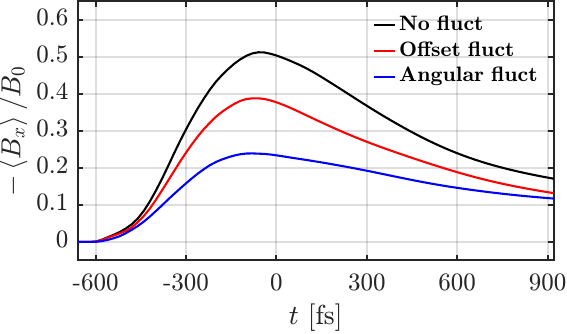}
    \caption{\rc{Impact of laser pointing stability on the generation of the axial magnetic field. The black curve is for the case without any fluctuations, using the setup shown in \cref{bx3d&r}. The red curve is for the case with beam offset fluctuations and the blue curve is for the case with with angular fluctuations. In each simulation, the axial magnetic field is averaged over a rectangular volume with$x\in[-19.9,-2.0]~\micron$ and $y$, $z\in[-8.0,8.0]~\micron$. Note that the laser beams leave the simulation domain at $t=0~\rm fs$.}}
    \label{cmp_jd}
\end{figure}

\rc{To examine the role of beam offset fluctuations, we introduce random displacements $\Delta x$ and $\Delta y$ for each beam. We change the position of a given beam hitting the target to $(x + \Delta x, y + \Delta y)$, where $x$ and $y$ are the locations prescribed by our setup. The range for the fluctuations is set to $|\Delta x| \leq 4~\micron$ and $|\Delta y| \leq 4~\micron$, whereas the original offset for each beam is $f_0 = \sqrt{x^2 + y^2} = 11~\micron$. Note that we keep the angles of incidence unchanged while varying the displacement. \Cref{cmp_jd} provides a comparison of the plasma magnetic field in the simulation with fluctuations and the plasma magnetic field in the simulation without fluctuations. Despite significant fluctuations ($\Delta f_0 / f_0 \approx 50\%$) the variation in the spatially averaged magnetic field is only around 10\%. The characteristic time scale at $t > 0$ remains unaffected, with the kT-scale field persisting over a ps.}

\rc{To examine the role of angular fluctuations, we introduce random variations $\Delta \theta$ and $\Delta \varphi$ for each beam. We adjust the angles for a given beam to $\theta + \Delta \theta$ and $\varphi + \Delta \varphi$, where $\theta = 0.27 \pi$ and $\varphi = 0.28 \pi$ are the convergence and twist angles prescribed by our setup. The range for the fluctuations is set to $|\Delta \theta| \leq 0.1 \pi$ and $|\Delta \varphi| \leq 0.1 \pi$. Note that we keep the beam displacement the same as in the original simulation. \Cref{cmp_jd} compares the plasma magnetic field in the simulation with fluctuations to the original result. The change in the spatially averaged magnetic field is about 50\%. Such sensitivity is not unexpected because we are varying the twist angle, which, as shown in \cref{Sec-4A}, is key to magnetic field generation. Despite the reduction in field strength, the characteristic time scale at $t > 0$ remains unaffected, with the kT-scale field again persisting over a ps.}

\rc{Our analysis suggests that angular fluctuations may have a more significant impact than displacement fluctuations. Given that the amplitude of the angular fluctuations in our simulation is approximately $18^{\circ}$, the obtained result demonstrates the robustness of our setup. In actual experiments, both types of fluctuations are likely to be present. To fully understand their combined impact, it is essential to consider the specifics of the laser architecture and the experimental setup.}

\rc{\subsection{Impact of density profile at the laser-irradiated surface}}

\rc{After examining the impact of various laser beam parameters, we now shift our focus to the influence of the target shape, specifically the density profile at the laser-irradiated surface. In all simulations presented in this paper, we used a solid target with an exponential preplasma, with parameters detailed in \Cref{LPI par}. This choice is motivated by the fact that high-intensity kJ-class laser systems inherently have a prepulse that inevitably causes target ablation and generates a density gradient at the front (laser-irradiated) surface. For comparison, in our previous work~\cite{Shi2023}, we used a solid target with initially intact nanowires to enhance laser absorption. A sufficiently thick preplasma can achieve this effect as well.}

\rc{To examine the impact of density profile at the laser-irradiated surface, we have performed two additional simulations: one with initially intact nanowires and one with an extended lower density shelf. The setup for the simulation with nanowires is similar to that used in our previous work~\cite{Shi2023}. The wires are $5~\micron$ long and $0.4~\micron$ thick. Their electron density is the same as that in the target ($n_e = 50 n_c$). The spacing between the wires is $2~\micron$. All other parameters are the same as in \Cref{LPI par}. In the simulation with the shelf, we replaced the exponential density profile at $x<0$ with a $5~\micron$ shelf whose density is $12.5 n_c$. The temporal evolution of the spatially-averaged longitudinal plasma magnetic field in these simulations is presented in \cref{Phs_contr}(b), alongside the curve for the original simulation with the exponential density profile. }

\rc{Although the general trend is similar for all three curves, there are several important conclusions that we can draw from \cref{Phs_contr}(b). First, we note that the nanowires used in our previous work~\cite{Shi2023} are not a must for the magnetic field generation. Furthermore, the exponential density profile and the shelf  generate a stronger field, with an improvement of nearly 10\% according to the simulation data. We also find that the target with the density shelf sustains the field for a longer period of time. At $t=900$~fs, its field strength is almost twice that of a target with the initially intact nanowires. These changes are likely associated with energy absorption and hot electron generation (e.g. see [\onlinecite{arefiev.pop.2016}], [\onlinecite{sorokovikova.prl.2016}], [\onlinecite{Peebles_2017}]), but a dedicated study is required to clearly identify the mechanisms at play. If achieving a low-density shelf is the goal, one might consider preheating the target in a controlled fashion~\cite{Peebles_2017} rather than relying on a prepulse to achieve this.}

\bigskip
\section{\label{Sec-5} Comparison with other schemes using CP or LG beams} 

In previous work on the IFE, CP Gaussian beams were among the first to be studied. For a CP laser beam, each photon carries a maximum SAM of $\sigma_z \hbar$, where $\sigma_z = + 1$ for right-hand 
and $\sigma_z = - 1$ for left-hand polarization.
Later, vortex beams such as LG beams that are known to carry OAM, were also used to study the IFE. For LG beams, each photon carries an OAM of $(\sigma_z + l)\hbar$, where $l$ is the twist index.
It is valuable to compare the magnetic field generation in our multi-beam scheme with the magnetic field generation via the IFE using CP or LG beams.
According to the model used in previous works~\cite{Haines2001, Ali2010, Longman2021} and assuming that the laser absorption rate and the electron density are constant in space and time, the following expression for the axial magnetic field can be obtained~(see Ref.~\onlinecite{Longman2021} for more details)
\begin{eqnarray}\label{B_z}
    B_z&=&\frac{2.25\eta_{\rm abs}I_0\tau}{ en_e\omega w_0^2}|\psi_l|^2\times\left[l\left(\frac{|l|w_0^2}{r^2}-2\right) \right. \nonumber\\
    & &\left. +\sigma_z\left(\frac{|l|^2w_0^2}{r^2}-4|l|+\frac{4r^2}{w_0^2}-2\right)\right]. 
\end{eqnarray} 
We have used a Gaussian temporal pulse shape, with $\tau$ being the temporal full width at half maximum (FWHM). In Eq.~(\ref{B_z}), $n_e$ is the electron density, $\eta_{\rm abs}$ is the laser energy absorption fraction per unit length, $I_0$ is the peak intensity and $w_0$ is the beam waist. 
Using reasonable assumptions, Longman~\cite{Longman2021} obtains the absorption rate as $\eta_{\rm abs}=n_e/(2n_cc\tau)$, then estimates the peak axial magnetic field strength of the $|\sigma_z|=1$, $|l|=0$ and $|\sigma_z|=0$, $|l|=1$ modes for $\tau\gtrsim100~\rm fs$ and $n_e\gtrsim0.01~n_c$
as $|B|_{\rm max}\approx (P~[{\rm TW}]\lambda^3~[\micron]/w_0^4~[\micron])~[10~\rm{kT}] $, where $P=0.75\pi I_0w_0^2$ is the laser power and $\eta_{\rm abs}\approx 0.2~\rm mm^{-1}$ is the value for the underdense plasma. The axial magnetic field $|B|_{\rm max}$ shows a fourth power dependence on the laser beam waist $w_0$. Meanwhile, under our high density plasma conditions, the absorption rate is expected to reach $\eta_{\rm abs}\simeq\eta_0\approx 32.3~\rm mm^{-1}$. Despite having a high absorption rate in our scenario, the total absorption may be reduced due to a limited interaction length of the laser absorption process caused by the presence of the high density plasma that is opaque to the laser beams. 

\begin{figure}[htb]
    \centering
    \includegraphics[width=0.9\columnwidth]{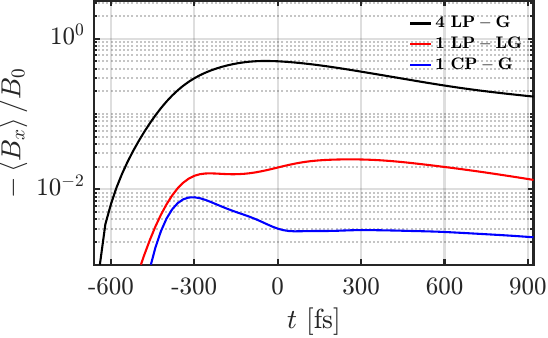}
    \caption{Average axial magnetic field within a rectangular volume of length $a=17.9~\micron$ ($x\in[-19.9,-2.0]~\micron$), width and height $b$, $c=16~\micron$ ($y$, $z\in[-8.0,8.0]~\micron$) as a function of time, where $t=0~\rm fs$ is defined as the moment when the lasers leave the simulation space completely. The black, red and blue solid lines represent the cases of four LP Gaussian laser beams, single LP LG mode laser, and single CP Gaussian mode laser, respectively.} 
    \label{Driv_contr}
\end{figure}

To make a quantitative comparison, we ran two 3D PIC simulations: one with a single CP Gaussian beam and another with a single LP LG beam. 
In these simulations, we set the laser parameters such that the laser energy is equal to the laser energy in the four beams from \cref{LPI par}. Additionally, we choose the beam waist radius to be equal to the beam offset ($w_0=f_0$) in \cref{Scheme} (b). This choice was made to ensure the transverse extension of the generated magnetic field matches that for the four-beam setup. Note that previous studies indicated that different laser modes with different values of $l$ and $\sigma_z$ produce radial magnetic field distributions with the intense region confined within $1.5w_0$, as shown in Ref.~\onlinecite{Longman2021}. 
It is worth pointing out that the magnetic field has a similar radial distribution for both the CP Gaussian beam ($|\sigma_z|=1$, $|l|=0$) and the LP LG beam ($|\sigma_z|=0$, $|l|=1$). 
Inserting these parameters into the expression of $|B|_{\rm max}$ above, we can estimate the axial magnetic field strength to be about 0.4 kT.  
In \cref{Driv_contr}, the logarithmic plots show the time evolution of the spatially averaged axial magnetic field for the three schemes mentioned here. The black curve represents the four LP Gaussian beams scenario. The red and blue curves represent the cases of a single LP LG and single CP Gaussian beam, respectively. It can be seen that the latter two have magnetic field strengths about two orders of magnitude lower than the former. In our simulations, the single CP Gaussian beam performs less effectively, while our multi-beam scheme shows distinct advantages under the same energy condition. 

\section {\label{Sec-6}Summary and discussion}

In summary, we have presented a comprehensive computational study of a setup where a strong axial magnetic field is generated in an interaction of multiple conventional laser beams with a target. The magnetic field strength  in the considered setup reaches 10 kT, with the strong field occupying tens of thousands of cubic microns. The field persists on a ps time scale. 
Although a CP Gaussian laser beam and an LG laser beam can carry intrinsic AM, native beams at current high-power laser systems are not CP or LG beams, but rather conventional LP beams.
Moreover, the design of kJ-level PW-class laser systems is such that they are composed of multiple LP beams.
This consideration motivated the development of our scheme that uses four laser beams with a twist in the pointing direction (shown in \cref{Scheme}) to achieve net OAM. 

The key role of the AM carried by these laser beams in initiating the magnetic field generation process has been established. The simulations examine various aspects, including the distribution of the axial magnetic field, azimuthal current, and electron and ion OAM densities. Dependencies on factors such as twist angle, polarization direction, delay, phase, and target structures have also been extensively studied. Our study has validated the effectiveness of this approach under various conditions, confirming its robustness and practical feasibility. In particular, the twist angle of the laser pulse emerges as a critical driver for maintaining the azimuthal plasma current that maintains the orientation of the magnetic field. 
Furthermore, the PIC simulations and supporting theory indicate that the twist angle serves as a convenient control knob for adjusting the direction and magnitude of the axial magnetic field. 

When compared to other schemes using CP or LG beams, the multi-beam configuration has several advantages. 
In particular, our approach requires only an LP Gaussian beam configuration, making it suitable for advanced high-power, high-intensity multi-beam laser systems such as the upcoming major upgrade,  SG-\Rmnum{2} UP. Despite the challenges associated with pointing directions, the results underscore the feasibility of achieving a strong and sustained axial magnetic field using thoughtfully designed multi-beam setups.
These studies contribute significantly to the understanding of laser-plasma interactions and expand the capabilities of high-power laser systems. 
The results may provide new opportunities to study the kT-scale magnetic field where the magnetic field energy density is greater than $10^{11}~\rm{J\cdot m}^{-3}$ which is usually the baseline in HED science. Other studies of strong magnetic field physics can also be expected.

\begin{acknowledgments}
Y. S. acknowledges the support by the National Natural Science Foundation of China (Grant No.~12322513), USTC Research Funds of the Double First-Class Initiative, CAS Project for Young Scientists in Basic Research (Grant No.~YSBR060). A. A. was supported by the Office of Fusion Energy Sciences under Award Numbers DE-SC0023423. 
Simulations were performed with EPOCH (developed under UK EPSRC Grants EP/G054950/1, EP/G056803/1, EP/G055165/1, and EP/M022463/1). The computational center of USTC and Hefei Advanced Computing Center are acknowledged for computational support.
\end{acknowledgments}

\section*{Data Availability Statement}

Data available on request from the authors. The data that support the findings of this study are available from the corresponding author upon reasonable request.



%

\end{document}